\documentstyle[pre,aps]{revtex}

\title{Dissipative Particle Dynamics with Energy Conservation:
Heat Conduction}

\author{Marisol Ripoll$^{(1)}$
\footnote{e-mail: mripoll@fisfun.uned.es, pep@fisfun.uned.es, ernst@phys.uu.nl}, 
Pep Espa\~{n}ol $^{(1)}$, and 
Matthieu H. Ernst$^{(2)}$}

\address{$^{(1)}$Dep. F\'{\i}sica Fundamental, UNED,
Apt. 60141, 28080 Madrid, Spain}

\address {$^{(2)}$Institute for Theoretical Physics,
Utrecht University,
Princetonplein 5, P.O. Box 80.006, 3508 TA Utrecht,
The Netherlands}

\begin{document}
\maketitle
\begin{abstract}  
We study by means of numerical
simulations the model of dissipative particle dynamics with energy
conservation for the simple case of thermal conduction. It is shown
that the model displays correct equilibrium fluctuations and
reproduces Fourier law.  The connection between ``mesoscopic
coarse-graining'' and ``resolution'' is clarified.
\end{abstract}

\section{Introduction}	
Mesoscopic approaches to the study of complex fluids are receiving a
great deal of attention in view of their potential to address
problems with disparate time-scales. The method of dissipative
particle dynamics introduced originally as an hybrid between lattice
gas and molecular dynamics simulations\cite{hoo92}, has proven to be a
flexible and powerful tool in the study of complex fluids as colloidal
suspensions\cite{koe93,boe97,boe97b}, porous flow\cite{hoo92}, polymer
suspensions\cite{sch95}, and multicomponent flows\cite{cov97}.

DPD models a fluid as a collection of point particles that are
interpreted as representing the center of mass of a mesoscopic cluster
of the atoms or molecules that constitute the fluid. By formulating
physically motivated interactions (that is, dissipative forces that
conserve momentum\cite{hoo92,esp95}) between these ``lumps'' of fluid,
it is assured that the macroscopic behavior of the system is
hydrodynamic\cite{esp95a,mar97}. Most importantly, a random
interaction is introduced in order to account for thermal fluctuations
that occur at the mesoscopic level and are responsible for Brownian
effects appearing in complex fluids.

One of the drawbacks of classic DPD is that the total energy of the
system is not conserved because the forces are velocity
dependent. This has been recently remedied simultaneously and
independently in Refs.\cite{esp97con,bon97}, where an internal energy
variable and a temperature is introduced for each particle. The
mechanical energy that is dissipated due to the velocity dependent
forces is transformed into internal energy of the particles. This
viscous heating process is accompanied by a thermal conduction process
where exchange of internal energy between neighboring particles occurs
when differences of temperatures exist. The new DPD algorithm with
energy conservation opens up the possibility of studying not only 
rheological aspects of complex fluids but also thermal issues. It is
therefore important to validate the technique in simple situations for
which the dynamical aspects are well-known, either analytically or by
means of other numerical simulations techniques.

As a first step towards the understanding of the behavior of DPD with
energy conservation we focus in this paper on the simplest problem of
conduction in a quiescent fluid (or a solid). This is a particular
case of the model introduced in Refs.\cite{esp97con,bon97}. The
particles are at rest and located randomly and represent portions of
material that can interchange energy. This simple model can be
regarded as a way of solving a fluctuating heat conduction
equation. In section 2 we review the model that was introduced in
Refs.\cite{esp97con,bon97}, for the case that only heat conduction
takes place.  In section 3 we further specify the model functions. In
section 4 we present simulation results that show the validity of the
simulation method by comparing the numerical results with the
theoretical results. Finally, we end up with a section of conclusions.

\section{The Model}
We have introduced in Ref.\cite{esp97con} a mesoscopic model for
describing a fluctuating viscous and thermally conducting fluid. Here
we apply this model to a quiescent fluid (or solid) within a box. The
system is represented by a set of $N$ particles of fixed positions
${\bf r}_i$ which are distributed at random within the box.  These
particles are understood as thermodynamic subsystems of the whole
system, that is, small portions of material that have a sufficiently
large number of degrees of freedom. To each particle we associate an
internal energy variable $\epsilon_i$ an entropy function
$s(\epsilon_i)$, and a temperature $T_i^{-1}=\partial s_i/\partial
\epsilon_i$. Each particle represents a mesoscopic portion of the
material and its internal energy content is subject to thermal
fluctuations, which in a continuum theory would be described by a
random heat flux.\cite{lan59} Differences in temperatures between
neighboring particles will cause a transfer of internal energy.
Therefore, the following stochastic differential equation for the
internal energy of each particle is postulated

\begin{equation}
d\epsilon_i=  \sum_j \omega(r_{ij})
\kappa_{ij}\left(\frac{1}{T_i}-\frac{1}{T_j}\right)dt+
\sum_j (2k_B\omega(r_{ij}) \kappa_{ij})^{1/2}d W^\epsilon_{ij}
\label{sdeok}
\end{equation}
The first term in the r.h.s. is deterministic and specifies that a
temperature difference causes exchange of energy. The
second term is stochastic and takes into account thermally induced
fluctuations in each particle caused by a random interchange of energy
between particles.  The notation is as follows.  $k_B$ is Boltzmann's
constant, $r_{ij}=|{\bf r}_i-{\bf r}_j|$ is the distance between
particles $i,j$ and the dimensionless weight function $\omega(r)$
determines the range of influence between particles. The parameter
$\kappa_{ij}$ governs the overall amplitude of the deterministic and
random parts. It depends in general on the state variables of
particles $i,j$ and is symmetric under particle interchange. The
random ``heat flux'' from particle $i$ to $j$ is expressed in terms of
the increments of the Wiener process $dW^\epsilon_{ij}$.  The
increments of the Wiener process are antisymmetric under particle
interchange $dW^\epsilon_{ij}=-dW^\epsilon_{ji}$ in such a way that
the total internal energy of the system governed by
Eqn. (\ref{sdeok}) is exactly conserved $\frac{d}{dt}\sum_i
\epsilon_i=0$. Because $\kappa_{ij}$ might depend in general on
$\epsilon_i,\epsilon_j$ we need to provide a stochastic interpretation
of Eqn. (\ref{sdeok}), as the noise is multiplicative. We choose It\^o interpretation.
It has been shown that, if $\partial_{ij}\kappa_{ij}=0$\footnote{This
implies that if $\kappa_{ij}$ depends only on $\epsilon_i,\epsilon_j$
then, necessarily, it will be a function of
$\epsilon_i+\epsilon_j$. This can be proved by considering a Taylor
expansion of a function of two variables and using the fact that, at
each point, both partial derivatives coincide.}, 
then Eqn. (\ref{sdeok}) is mathematically equivalent to a
Fokker-Planck equation which has as equilibrium solution 
\begin{equation}
\rho_{eq}(\epsilon) 
\propto \exp\left\{k_B^{-1}\sum_i s(\epsilon_i) \right\}
{\cal P}(\sum_k\epsilon_k)
\label{equilcon}
\end{equation}
where ${\cal
P}(\sum_k\epsilon_k)$ is a function of the total internal energy which
is determined by the initial distribution of total
energy. Particular examples are given by the canonical\cite{bon97} and
microcanonical  ensembles 
\begin{eqnarray}
\rho_{mic}(\epsilon_1,\ldots,\epsilon_N)
&=&
\frac{1}{\Omega(E_0,N)} \exp\left\{k_B^{-1}\sum_i s(\epsilon_i) \right\}
\delta(\sum_i\epsilon_i-E_0)
\nonumber\\
\rho_{can}(\epsilon_1,\ldots,\epsilon_N)
&=&
\frac{1}{Z(\beta,N)} \exp\left\{k_B^{-1}\sum_i s(\epsilon_i)-\beta\epsilon_i \right\}
\label{miccan}
\end{eqnarray}
Here, the factors $\Omega(E_0,N)$ and $Z(\beta,N)$ are obtained by
normalization.

\section{Perfect solid and Fourier law}
In this section we further specify the model by providing the
undetermined functions $T(\epsilon)$, $\kappa(\epsilon+\epsilon')$ and
$\omega(r)$ appearing in Eqn. (\ref{sdeok}).

The simplest possible form for the equation of state $T(\epsilon)$ is
that of a perfect solid, which is a very good approximation for
metals. The perfect solid has the following equation of state

\begin{equation}
T(\epsilon)=\frac{\epsilon}{C_v}
\label{ecstate}
\end{equation}
where the heat capacity of the mesoscopic particles $C_v$ is a
constant independent of the energy. It is an extensive property that
depends on the ``size'' of the particles. For mesoscopic particles,
the dimensionless heat capacity $\alpha \equiv C_v/k_B\gg 1$. The
entropy $s(\epsilon)$ is obtained by integrating the temperature and,
apart from irrelevant constant additive terms, it is given by
$s(\epsilon)= C_v \ln \epsilon$.

Now, let us assume the following functional form for $\kappa_{ij}$,

\begin{equation}
\kappa_{ij}=\frac{C_v\tilde{\kappa}}{\lambda^2}
T^2\left(\frac{\epsilon_i+\epsilon_j}{2}\right)
=\frac{C_v\tilde{\kappa}}{4\lambda^2}
\left(T_i+T_j\right)^2
\label{kfou}
\end{equation}
where $\lambda$ is the average distance between particles (this is
$\lambda=n^{-1/3}$ and $n$ is the average density number) and
$\tilde{\kappa}$ has dimensions of a diffusion coefficient.
In the last equality, Eqn. (\ref{ecstate}) has been used.

Substitution of Eqn. (\ref{kfou}) into
the dynamical equation for the energy leads to the following
equation for the temperature of a particle
\begin{equation}
dT_i =
\frac{\tilde{\kappa}}{\lambda^2} \sum_j\omega(r_{ij})
\frac{(T_i+T_j)^2}{4T_iT_j}
(T_j-T_i)dt+
\left(\frac{2k_B\tilde{\kappa}}{C_v\lambda^2}\right)^{1/2}
\sum_j\omega^{1/2}(r_{ij})\frac{T_i+T_j}{2}d W^\epsilon_{ij}
\label{T1}
\end{equation}
The factor $\frac{(T_i+T_j)^2}{4T_iT_j}$ should be very close to 1 if
the temperature differences between particles within an action sphere
are small.  We see, then, that for the perfect solid by a proper
selection of $\kappa_{ij}$ the form of the postulated SDE for the
energy leads to a discrete version of the Fourier law of heat
conduction. Note that the factor $\lambda^2$ in the first term of the
left hand side of Eqn. (\ref{T1}) can be understood as the ``lattice
spacing'' squared that would arise in a finite difference
discretization of the Laplacian in the heat conduction equation.  In a
sense, Eqn. (\ref{T1}) can be understood as a finite difference
discretization of the macroscopic heat equation on a {\em random}
lattice which, in addition, has the correct form for the thermal
fluctuations.

We will assume that the weight
function $\omega(r)$ has the form of the Lucy 
weight function\cite{luc77}, which is a smooth bell-shaped function with
continuous derivatives. It is given by

\begin{equation}
\omega(r) = \frac{105}{16\pi s^3}\left(1+3\frac{r}{r_c}\right)
\left(1-\frac{r}{r_c}\right)^3
\label{omegalucy}
\end{equation}
where $r_c$ is range of interaction between particle and
$s=r_c/\lambda$ is the {\em overlapping coefficient} which gives,
essentially, the number of neighbors of a given particle. We have
normalized the weight function as\cite{hoo92}

\begin{equation}
\int d^d{\bf r} \omega(r) = \frac{1}{n}
\label{orignorm}
\end{equation}

\section{Simulation results}
In this section we present some results obtained from numerical
simulation of Eqn. (\ref{T1}). The physical system is assumed to be in
a three dimensional cubic box of edge length $L$. Initially the box is
seeded with $N$ mesoscopic particles located at random. The density is
then $n=N/L^3$ and the typical interparticle distance is
$\lambda=n^{-1/3}$.  Periodic boundary conditions are assumed in the
$y,z$ directions and in the $x$ direction we assume either periodic
boundary conditions when considering equilibrium situations or we
impose the temperature at $x=-L/2$ and $x=L/2$. This is achieved by
considering two extra layers of particles in each boundary in the $x$
direction filled with particles that are at a constant
temperature. These layers act, therefore, as {\em thermal baths}.
They have a width as large as $r_c$ in order that any particle within
the system interacts with the same number of particles.

The basic model parameters are the following ones: $\tilde{\kappa}$,
$N$, $L$, $r_c$, $C_v$, $T_0$. Here, $T_0$ is a reference temperature,
that of equilibrium or, in the case of nonequilibrium situations, that
of the colder thermal bath. We work in reduced units such that $L$ is
the unit of length, $L^2/\tilde{\kappa}$ is unit of time, $T_0$ is the
unit of temperature, and $k_B$ is taken as the unit of entropy or heat
capacity. 

We introduce the dimensionless heat capacity of mesoscopic particles
$\alpha=C_v/k_B$ and the dimensionless heat capacity of the total
simulated sample, which is $C_L=N\alpha$ because heat capacity is an
extensive property. On the other hand, $C_L =L^3c_v/k_B$, where $c_v$
is the heat capacity per unit volume, which is a material constant.
For example, $c_{v}=3.44 \times 10^{6}J/Km^3$ for copper at room
temperature. Therefore, by fixing $C_L$ we are fixing the actual
volume of the sample. For copper, a value $C_L=10^8$ implies a
submicron sample size of $L=C_L^{1/3} \times 1.6 10^{-10}m\sim
0.1\mu$. In most of the simulations we use $C_L=10^8$ and this fixes
the value of $\alpha$ depending on the number of particles in the
system. Note that by fixing $C_L$, large $N$ implies small $\alpha$,
in accordance with the idea that the heat capacity of the mesoscopic
particles is proportional to its typical size. We note that the
overall intensity of the noise in (\ref{T1}) is proportional to
$(k_B/C_v)^{1/2}=\alpha^{-1/2}$ and, therefore, inversely proportional
to the square root of the volume of the particles.\cite{esp97sde} The
higher the number density of mesoscopic particles used to discretize a
given sample, the smaller is the size of such particles and,
therefore, the larger is the thermal noise.

Eqns. (\ref{T1}) are solved with a conventional Euler method.  For
values of $\alpha<10$, the stochastic term produces from time
to time a temperature which is negative for some particle. This has
disastrous consequences because the factor accompanying $(T_i-T_j)$ in
the deterministic term of (\ref{T1}) is negative and produces a flow
of energy from the cold particles to the hot particles. When this
happens the system becomes unstable. However, this never occurs for
sufficiently large values of $\alpha$ as the ones we typically use and
no instabilities are observed.

In order to check that the code performs properly, we first run a
simulation with fully periodic boundary conditions. Initially, all the
particles are assumed to have the same temperature $T=1$. After some
equilibration period the system reaches equilibrium. The energy is
exactly conserved (no round-off errors) because we use a Verlet
list\cite{all}, in such a way that what is gained by a particle is
exactly what is lost by the rest of particles.

\begin{figure}[ht]
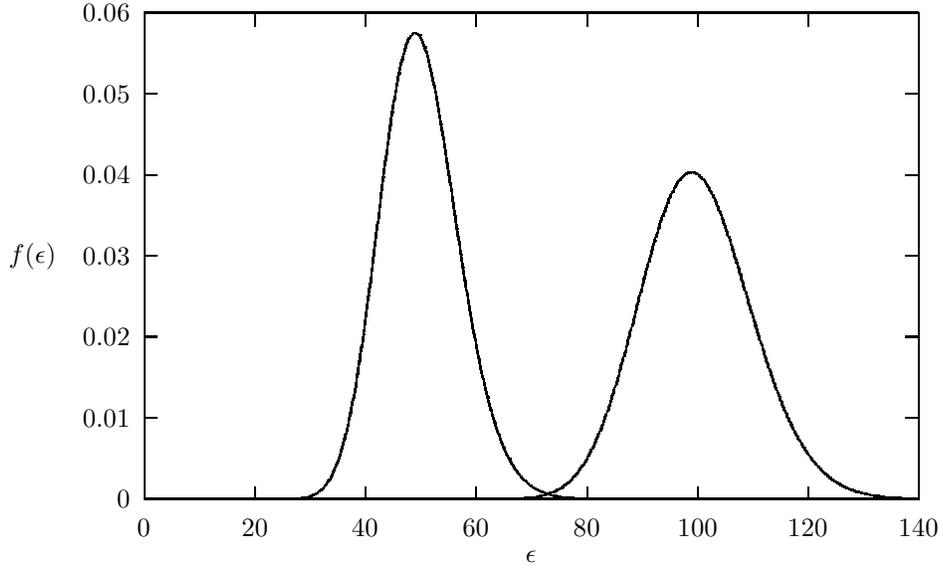

\begin{center}
\setlength{\unitlength}{0.240900pt}
\ifx\plotpoint\undefined\newsavebox{\plotpoint}\fi
\sbox{\plotpoint}{\rule[-0.200pt]{0.400pt}{0.400pt}}%


\caption{The probability density distribution $f(\epsilon)$ of the
energy of a given particle. Two simulations with $\alpha=50$ (left)
and $\alpha=100$ (right) are presented. Other parameters are $N=1000$,
$s=1.5$, $T_{eq}=1$. Also shown, although not visible because they are
perfectly superimposed, are the theoretical predictions for the
microcanonical and canonical distributions for the given values of
$\alpha$.}
\label{fig1}
\end{center}
\end{figure}

In Fig. \ref{fig1} we show the probability density distribution of the
energy of a particle as obtained from simulation of Eqn. (\ref{sdeok}) for
two different values of $\alpha=C_v/k_B$, the dimensionless heat
capacity of the mesoscopic particles, for a system of $N=1000$. The
larger $\alpha$ the larger the average value of the energy and
variance. However, the relative fluctuations are smaller,
proportional to $N^{-1/2}$. Also shown in Fig. \ref{fig1} are the
canonical $f_{\rm can}(\epsilon)$ and microcanonical $f_{\rm
mic}(\epsilon)$ theoretical results for the given values of
$\alpha$. The analytical expressions can be obtained from (\ref{miccan}) 
and are given by 

\begin{eqnarray}
f_{\rm can}(\epsilon) &=&
\frac{\beta}{\Gamma(\alpha+1)}\left(\beta\epsilon\right)^\alpha
\exp\{-\beta\epsilon\}
\nonumber\\
f_{\rm mic}(\epsilon) &=&
\frac{1}{M}\frac{1}{E_0}\left(\frac{\epsilon}{E_0}\right)^{\alpha}
\left(1-\frac{\epsilon}{E_0}\right)^{(N-1)(\alpha+1)-1}
\label{fcan}
\end{eqnarray}
where $\Gamma(x)$ is the gamma function. The parameter $\beta$ can be
explicitly computed for a perfect solid by requiring that the average
energy of a particle is $E_0/N$, where $E_0$ is the total energy of
the isolated system. The result is $\beta= N(\alpha+1)/E_0$. In the
thermodynamic limit the canonical and microcanonical distributions are
identical. In practice, they are indistinguishable for values $N\alpha>50$.
The perfect coincidence of the theoretical and simulation results in Fig. \ref{fig1}
gives confidence on the implementation of the numerical code used.

Next, we compute the thermal diffusivity by a macroscopic
measurement. Two heat baths are constructed with temperatures $T_{\rm
cold}$ and $T_{\rm hot}$ in such a way that a gradient $(T_{\rm
hot}-T_{\rm cold})/L$ is applied. A steady state is reached
for sufficiently long times. Fig. \ref{figgrad} shows the average of
the temperature once the steady state has been reached for different
number of particles for the case that $T_{\rm cold}=1$, $T_{\rm
hot}=2$ and the overlapping coefficient is $s=1.5$. For $N=100$ the
spatial noise is considerable. Fluctuations in this graph are not due
to statistical noise but to the spatial inhomogeneities that appear
for small number of particles and they would smear out by averaging
over different realizations of the initial spatial configuration of
the particles. For larger number of particles a smooth linear profile
is achieved, in accordance with our expectation that the discrete
model reproduces Fourier law. The temperature field $T(x)$ is obtained
by binning the $x$ axis and averaging the temperatures of the particles
within each bin of volume $\Delta x L^2$.

We have performed a series of simulations with different temperature
gradients and have computed the macroscopic heat flux in the steady
state. Two different ways of computing the macroscopic heat flux have
been used. In the first case, the macroscopic heat flux is obtained by
computing at each time step the energy gained by the cold bath (which
is equal to the energy lost by the hot bath) and dividing it by the
time step, this is $\Delta E/\Delta t$. The second way follows from
the microscopic definition of the heat flux as obtained from
projection operator methods or from kinetic theory\cite{preprint}
\begin{equation}
{\bf Q} \equiv 
\sum_{ij}\omega(r_{ij})\kappa_{ij}
\left[\frac{1}{T_i}-\frac{1}{T_j}\right]
\frac{{\bf r}_{ij}}{2}
\label{Qtotal}
\end{equation}
All transport fluxes are caused by collisional transfer. As the particles
do no move, there are no kinetic fluxes.
This second form produces better
results because it involves a very large number of pairs of particles
and it is therefore adopted here. 

In Fig. \ref{figq} we plot the
macroscopic heat flux in terms of the applied temperature gradient. A
linear dependence is obtained, which allows to extract the thermal
diffusivity from the slope.

When this experiment is performed for different number densities and
different overlapping we can compare with the results from kinetic
theory as developed in Ref.\cite{preprint}. Fig. \ref{ds2} shows the
value of the macroscopic thermal diffusivity as a function of the
overlapping $s$. Also shown is the kinetic theory prediction for the
thermal diffusivity\cite{preprint}, reading,
\begin{equation}
D= \frac{s^2}{24}\tilde{\kappa}
\label{Dtheory}
\end{equation}
We observe that the theoretical prediction improves for large overlapping. 

The kinetic theory result (\ref{Dtheory}) has been obtained through
the choices Eqns. (\ref{ecstate}) and (\ref{kfou}) and shows no
dependence on the density. This is a consequence of introducing the
factor $\lambda ^{-2}$ in Eqn. (\ref{kfou}). We verify in Fig.
\ref{dn} that simulations performed at different number density, with
the remaining parameters $s,C_L$ held fixed, show that the thermal
diffusivity $D$ is indeed independent of $n$, as predicted in
Eqn. (\ref{Dtheory}).

\section{Conclusions}

In this paper we have applied the model introduced in
Ref.\cite{esp97con} to the case of heat conduction in a random
solid. We have chosen the equation of state $s(\epsilon_i)$ for a
perfect solid model and a particular model for $\kappa_{ij}$,
Eqn. (\ref{kfou}). It has been shown that the model has the correct
equilibrium properties and that it reproduces Fourier's law in
non-equilibrium situations. We have also corroborated, by means of
simulations, the kinetic theory predictions for this model which are
presented elsewhere. Kinetic theory predicts a thermal diffusivity
which depends only on the overlapping and not on the density. By
introducing the factor $\lambda^{-2}$ in the definition of
$\kappa_{ij}$ we have shown that it is possible to interpret $n$ not
as the density but as the {\em resolution} at which the physical
problem is being studied. A similar discussion about resolution was
presented for the case of DPD with no energy conservation in
Ref.\cite{esp97fpm}.

\section*{Acknowledgments}
\noindent
M.R. acknowledges the support of a F.P.I. grant from Ministerio
de Educaci\'on. P.E. has received partial support from 
DGICYT Project No PB94-0382 and by E.C. Contract ERB-CHRXCT-940546.

\begin{figure}[ht]
\begin{center}
\setlength{\unitlength}{0.240900pt}
\ifx\plotpoint\undefined\newsavebox{\plotpoint}\fi
\sbox{\plotpoint}{\rule[-0.200pt]{0.400pt}{0.400pt}}%
\begin{picture}(1500,900)(0,0)
\font\gnuplot=cmr10 at 10pt
\gnuplot
\sbox{\plotpoint}{\rule[-0.200pt]{0.400pt}{0.400pt}}%
\put(828.0,113.0){\rule[-0.200pt]{0.400pt}{184.048pt}}
\put(220.0,113.0){\rule[-0.200pt]{4.818pt}{0.400pt}}
\put(198,113){\makebox(0,0)[r]{1}}
\put(1416.0,113.0){\rule[-0.200pt]{4.818pt}{0.400pt}}
\put(220.0,189.0){\rule[-0.200pt]{4.818pt}{0.400pt}}
\put(198,189){\makebox(0,0)[r]{1.1}}
\put(1416.0,189.0){\rule[-0.200pt]{4.818pt}{0.400pt}}
\put(220.0,266.0){\rule[-0.200pt]{4.818pt}{0.400pt}}
\put(198,266){\makebox(0,0)[r]{1.2}}
\put(1416.0,266.0){\rule[-0.200pt]{4.818pt}{0.400pt}}
\put(220.0,342.0){\rule[-0.200pt]{4.818pt}{0.400pt}}
\put(198,342){\makebox(0,0)[r]{1.3}}
\put(1416.0,342.0){\rule[-0.200pt]{4.818pt}{0.400pt}}
\put(220.0,419.0){\rule[-0.200pt]{4.818pt}{0.400pt}}
\put(198,419){\makebox(0,0)[r]{1.4}}
\put(1416.0,419.0){\rule[-0.200pt]{4.818pt}{0.400pt}}
\put(220.0,495.0){\rule[-0.200pt]{4.818pt}{0.400pt}}
\put(198,495){\makebox(0,0)[r]{1.5}}
\put(1416.0,495.0){\rule[-0.200pt]{4.818pt}{0.400pt}}
\put(220.0,571.0){\rule[-0.200pt]{4.818pt}{0.400pt}}
\put(198,571){\makebox(0,0)[r]{1.6}}
\put(1416.0,571.0){\rule[-0.200pt]{4.818pt}{0.400pt}}
\put(220.0,648.0){\rule[-0.200pt]{4.818pt}{0.400pt}}
\put(198,648){\makebox(0,0)[r]{1.7}}
\put(1416.0,648.0){\rule[-0.200pt]{4.818pt}{0.400pt}}
\put(220.0,724.0){\rule[-0.200pt]{4.818pt}{0.400pt}}
\put(198,724){\makebox(0,0)[r]{1.8}}
\put(1416.0,724.0){\rule[-0.200pt]{4.818pt}{0.400pt}}
\put(220.0,801.0){\rule[-0.200pt]{4.818pt}{0.400pt}}
\put(198,801){\makebox(0,0)[r]{1.9}}
\put(1416.0,801.0){\rule[-0.200pt]{4.818pt}{0.400pt}}
\put(220.0,877.0){\rule[-0.200pt]{4.818pt}{0.400pt}}
\put(198,877){\makebox(0,0)[r]{2}}
\put(1416.0,877.0){\rule[-0.200pt]{4.818pt}{0.400pt}}
\put(220.0,113.0){\rule[-0.200pt]{0.400pt}{4.818pt}}
\put(220,68){\makebox(0,0){-0.5}}
\put(220.0,857.0){\rule[-0.200pt]{0.400pt}{4.818pt}}
\put(342.0,113.0){\rule[-0.200pt]{0.400pt}{4.818pt}}
\put(342,68){\makebox(0,0){-0.4}}
\put(342.0,857.0){\rule[-0.200pt]{0.400pt}{4.818pt}}
\put(463.0,113.0){\rule[-0.200pt]{0.400pt}{4.818pt}}
\put(463,68){\makebox(0,0){-0.3}}
\put(463.0,857.0){\rule[-0.200pt]{0.400pt}{4.818pt}}
\put(585.0,113.0){\rule[-0.200pt]{0.400pt}{4.818pt}}
\put(585,68){\makebox(0,0){-0.2}}
\put(585.0,857.0){\rule[-0.200pt]{0.400pt}{4.818pt}}
\put(706.0,113.0){\rule[-0.200pt]{0.400pt}{4.818pt}}
\put(706,68){\makebox(0,0){-0.1}}
\put(706.0,857.0){\rule[-0.200pt]{0.400pt}{4.818pt}}
\put(828.0,113.0){\rule[-0.200pt]{0.400pt}{4.818pt}}
\put(828,68){\makebox(0,0){0}}
\put(828.0,857.0){\rule[-0.200pt]{0.400pt}{4.818pt}}
\put(950.0,113.0){\rule[-0.200pt]{0.400pt}{4.818pt}}
\put(950,68){\makebox(0,0){0.1}}
\put(950.0,857.0){\rule[-0.200pt]{0.400pt}{4.818pt}}
\put(1071.0,113.0){\rule[-0.200pt]{0.400pt}{4.818pt}}
\put(1071,68){\makebox(0,0){0.2}}
\put(1071.0,857.0){\rule[-0.200pt]{0.400pt}{4.818pt}}
\put(1193.0,113.0){\rule[-0.200pt]{0.400pt}{4.818pt}}
\put(1193,68){\makebox(0,0){0.3}}
\put(1193.0,857.0){\rule[-0.200pt]{0.400pt}{4.818pt}}
\put(1314.0,113.0){\rule[-0.200pt]{0.400pt}{4.818pt}}
\put(1314,68){\makebox(0,0){0.4}}
\put(1314.0,857.0){\rule[-0.200pt]{0.400pt}{4.818pt}}
\put(1436.0,113.0){\rule[-0.200pt]{0.400pt}{4.818pt}}
\put(1436,68){\makebox(0,0){0.5}}
\put(1436.0,857.0){\rule[-0.200pt]{0.400pt}{4.818pt}}
\put(220.0,113.0){\rule[-0.200pt]{292.934pt}{0.400pt}}
\put(1436.0,113.0){\rule[-0.200pt]{0.400pt}{184.048pt}}
\put(220.0,877.0){\rule[-0.200pt]{292.934pt}{0.400pt}}
\put(45,495){\makebox(0,0){${T({x})}$}}
\put(828,23){\makebox(0,0){${x}$}}
\put(220.0,113.0){\rule[-0.200pt]{0.400pt}{184.048pt}}
\sbox{\plotpoint}{\rule[-0.400pt]{0.800pt}{0.800pt}}%
\put(244,139){\raisebox{-.8pt}{\makebox(0,0){${\bullet}$}}}
\put(293,149){\raisebox{-.8pt}{\makebox(0,0){${\bullet}$}}}
\put(342,203){\raisebox{-.8pt}{\makebox(0,0){${\bullet}$}}}
\put(390,231){\raisebox{-.8pt}{\makebox(0,0){${\bullet}$}}}
\put(439,253){\raisebox{-.8pt}{\makebox(0,0){${\bullet}$}}}
\put(488,272){\raisebox{-.8pt}{\makebox(0,0){${\bullet}$}}}
\put(536,305){\raisebox{-.8pt}{\makebox(0,0){${\bullet}$}}}
\put(585,329){\raisebox{-.8pt}{\makebox(0,0){${\bullet}$}}}
\put(633,360){\raisebox{-.8pt}{\makebox(0,0){${\bullet}$}}}
\put(682,390){\raisebox{-.8pt}{\makebox(0,0){${\bullet}$}}}
\put(731,437){\raisebox{-.8pt}{\makebox(0,0){${\bullet}$}}}
\put(779,477){\raisebox{-.8pt}{\makebox(0,0){${\bullet}$}}}
\put(828,493){\raisebox{-.8pt}{\makebox(0,0){${\bullet}$}}}
\put(877,534){\raisebox{-.8pt}{\makebox(0,0){${\bullet}$}}}
\put(925,542){\raisebox{-.8pt}{\makebox(0,0){${\bullet}$}}}
\put(974,576){\raisebox{-.8pt}{\makebox(0,0){${\bullet}$}}}
\put(1023,601){\raisebox{-.8pt}{\makebox(0,0){${\bullet}$}}}
\put(1071,633){\raisebox{-.8pt}{\makebox(0,0){${\bullet}$}}}
\put(1120,687){\raisebox{-.8pt}{\makebox(0,0){${\bullet}$}}}
\put(1168,713){\raisebox{-.8pt}{\makebox(0,0){${\bullet}$}}}
\put(1217,753){\raisebox{-.8pt}{\makebox(0,0){${\bullet}$}}}
\put(1266,785){\raisebox{-.8pt}{\makebox(0,0){${\bullet}$}}}
\put(1314,803){\raisebox{-.8pt}{\makebox(0,0){${\bullet}$}}}
\put(1363,832){\raisebox{-.8pt}{\makebox(0,0){${\bullet}$}}}
\put(1412,846){\raisebox{-.8pt}{\makebox(0,0){${\bullet}$}}}
\sbox{\plotpoint}{\rule[-0.200pt]{0.400pt}{0.400pt}}%
\put(244,134){\makebox(0,0){${\circ} $}}
\put(293,210){\makebox(0,0){${\circ} $}}
\put(342,209){\makebox(0,0){${\circ} $}}
\put(390,222){\makebox(0,0){${\circ} $}}
\put(439,196){\makebox(0,0){${\circ} $}}
\put(488,341){\makebox(0,0){${\circ} $}}
\put(536,303){\makebox(0,0){${\circ} $}}
\put(585,415){\makebox(0,0){${\circ} $}}
\put(633,219){\makebox(0,0){${\circ} $}}
\put(682,465){\makebox(0,0){${\circ} $}}
\put(731,494){\makebox(0,0){${\circ} $}}
\put(779,519){\makebox(0,0){${\circ} $}}
\put(828,557){\makebox(0,0){${\circ} $}}
\put(877,535){\makebox(0,0){${\circ} $}}
\put(925,676){\makebox(0,0){${\circ} $}}
\put(974,637){\makebox(0,0){${\circ} $}}
\put(1023,721){\makebox(0,0){${\circ} $}}
\put(1071,722){\makebox(0,0){${\circ} $}}
\put(1120,744){\makebox(0,0){${\circ} $}}
\put(1168,748){\makebox(0,0){${\circ} $}}
\put(1217,778){\makebox(0,0){${\circ} $}}
\put(1266,782){\makebox(0,0){${\circ} $}}
\put(1314,807){\makebox(0,0){${\circ} $}}
\put(1363,842){\makebox(0,0){${\circ} $}}
\put(1412,835){\makebox(0,0){${\circ} $}}
\put(244,128){\usebox{\plotpoint}}
\multiput(244.00,128.59)(0.758,0.488){13}{\rule{0.700pt}{0.117pt}}
\multiput(244.00,127.17)(10.547,8.000){2}{\rule{0.350pt}{0.400pt}}
\multiput(256.00,136.59)(0.874,0.485){11}{\rule{0.786pt}{0.117pt}}
\multiput(256.00,135.17)(10.369,7.000){2}{\rule{0.393pt}{0.400pt}}
\multiput(268.00,143.59)(0.758,0.488){13}{\rule{0.700pt}{0.117pt}}
\multiput(268.00,142.17)(10.547,8.000){2}{\rule{0.350pt}{0.400pt}}
\multiput(280.00,151.59)(0.798,0.485){11}{\rule{0.729pt}{0.117pt}}
\multiput(280.00,150.17)(9.488,7.000){2}{\rule{0.364pt}{0.400pt}}
\multiput(291.00,158.59)(0.874,0.485){11}{\rule{0.786pt}{0.117pt}}
\multiput(291.00,157.17)(10.369,7.000){2}{\rule{0.393pt}{0.400pt}}
\multiput(303.00,165.59)(0.758,0.488){13}{\rule{0.700pt}{0.117pt}}
\multiput(303.00,164.17)(10.547,8.000){2}{\rule{0.350pt}{0.400pt}}
\multiput(315.00,173.59)(0.874,0.485){11}{\rule{0.786pt}{0.117pt}}
\multiput(315.00,172.17)(10.369,7.000){2}{\rule{0.393pt}{0.400pt}}
\multiput(327.00,180.59)(0.758,0.488){13}{\rule{0.700pt}{0.117pt}}
\multiput(327.00,179.17)(10.547,8.000){2}{\rule{0.350pt}{0.400pt}}
\multiput(339.00,188.59)(0.798,0.485){11}{\rule{0.729pt}{0.117pt}}
\multiput(339.00,187.17)(9.488,7.000){2}{\rule{0.364pt}{0.400pt}}
\multiput(350.00,195.59)(0.874,0.485){11}{\rule{0.786pt}{0.117pt}}
\multiput(350.00,194.17)(10.369,7.000){2}{\rule{0.393pt}{0.400pt}}
\multiput(362.00,202.59)(0.758,0.488){13}{\rule{0.700pt}{0.117pt}}
\multiput(362.00,201.17)(10.547,8.000){2}{\rule{0.350pt}{0.400pt}}
\multiput(374.00,210.59)(0.874,0.485){11}{\rule{0.786pt}{0.117pt}}
\multiput(374.00,209.17)(10.369,7.000){2}{\rule{0.393pt}{0.400pt}}
\multiput(386.00,217.59)(0.758,0.488){13}{\rule{0.700pt}{0.117pt}}
\multiput(386.00,216.17)(10.547,8.000){2}{\rule{0.350pt}{0.400pt}}
\multiput(398.00,225.59)(0.798,0.485){11}{\rule{0.729pt}{0.117pt}}
\multiput(398.00,224.17)(9.488,7.000){2}{\rule{0.364pt}{0.400pt}}
\multiput(409.00,232.59)(0.874,0.485){11}{\rule{0.786pt}{0.117pt}}
\multiput(409.00,231.17)(10.369,7.000){2}{\rule{0.393pt}{0.400pt}}
\multiput(421.00,239.59)(0.758,0.488){13}{\rule{0.700pt}{0.117pt}}
\multiput(421.00,238.17)(10.547,8.000){2}{\rule{0.350pt}{0.400pt}}
\multiput(433.00,247.59)(0.874,0.485){11}{\rule{0.786pt}{0.117pt}}
\multiput(433.00,246.17)(10.369,7.000){2}{\rule{0.393pt}{0.400pt}}
\multiput(445.00,254.59)(0.758,0.488){13}{\rule{0.700pt}{0.117pt}}
\multiput(445.00,253.17)(10.547,8.000){2}{\rule{0.350pt}{0.400pt}}
\multiput(457.00,262.59)(0.798,0.485){11}{\rule{0.729pt}{0.117pt}}
\multiput(457.00,261.17)(9.488,7.000){2}{\rule{0.364pt}{0.400pt}}
\multiput(468.00,269.59)(0.874,0.485){11}{\rule{0.786pt}{0.117pt}}
\multiput(468.00,268.17)(10.369,7.000){2}{\rule{0.393pt}{0.400pt}}
\multiput(480.00,276.59)(0.758,0.488){13}{\rule{0.700pt}{0.117pt}}
\multiput(480.00,275.17)(10.547,8.000){2}{\rule{0.350pt}{0.400pt}}
\multiput(492.00,284.59)(0.874,0.485){11}{\rule{0.786pt}{0.117pt}}
\multiput(492.00,283.17)(10.369,7.000){2}{\rule{0.393pt}{0.400pt}}
\multiput(504.00,291.59)(0.758,0.488){13}{\rule{0.700pt}{0.117pt}}
\multiput(504.00,290.17)(10.547,8.000){2}{\rule{0.350pt}{0.400pt}}
\multiput(516.00,299.59)(0.798,0.485){11}{\rule{0.729pt}{0.117pt}}
\multiput(516.00,298.17)(9.488,7.000){2}{\rule{0.364pt}{0.400pt}}
\multiput(527.00,306.59)(0.874,0.485){11}{\rule{0.786pt}{0.117pt}}
\multiput(527.00,305.17)(10.369,7.000){2}{\rule{0.393pt}{0.400pt}}
\multiput(539.00,313.59)(0.758,0.488){13}{\rule{0.700pt}{0.117pt}}
\multiput(539.00,312.17)(10.547,8.000){2}{\rule{0.350pt}{0.400pt}}
\multiput(551.00,321.59)(0.874,0.485){11}{\rule{0.786pt}{0.117pt}}
\multiput(551.00,320.17)(10.369,7.000){2}{\rule{0.393pt}{0.400pt}}
\multiput(563.00,328.59)(0.692,0.488){13}{\rule{0.650pt}{0.117pt}}
\multiput(563.00,327.17)(9.651,8.000){2}{\rule{0.325pt}{0.400pt}}
\multiput(574.00,336.59)(0.874,0.485){11}{\rule{0.786pt}{0.117pt}}
\multiput(574.00,335.17)(10.369,7.000){2}{\rule{0.393pt}{0.400pt}}
\multiput(586.00,343.59)(0.758,0.488){13}{\rule{0.700pt}{0.117pt}}
\multiput(586.00,342.17)(10.547,8.000){2}{\rule{0.350pt}{0.400pt}}
\multiput(598.00,351.59)(0.874,0.485){11}{\rule{0.786pt}{0.117pt}}
\multiput(598.00,350.17)(10.369,7.000){2}{\rule{0.393pt}{0.400pt}}
\multiput(610.00,358.59)(0.874,0.485){11}{\rule{0.786pt}{0.117pt}}
\multiput(610.00,357.17)(10.369,7.000){2}{\rule{0.393pt}{0.400pt}}
\multiput(622.00,365.59)(0.692,0.488){13}{\rule{0.650pt}{0.117pt}}
\multiput(622.00,364.17)(9.651,8.000){2}{\rule{0.325pt}{0.400pt}}
\multiput(633.00,373.59)(0.874,0.485){11}{\rule{0.786pt}{0.117pt}}
\multiput(633.00,372.17)(10.369,7.000){2}{\rule{0.393pt}{0.400pt}}
\multiput(645.00,380.59)(0.758,0.488){13}{\rule{0.700pt}{0.117pt}}
\multiput(645.00,379.17)(10.547,8.000){2}{\rule{0.350pt}{0.400pt}}
\multiput(657.00,388.59)(0.874,0.485){11}{\rule{0.786pt}{0.117pt}}
\multiput(657.00,387.17)(10.369,7.000){2}{\rule{0.393pt}{0.400pt}}
\multiput(669.00,395.59)(0.874,0.485){11}{\rule{0.786pt}{0.117pt}}
\multiput(669.00,394.17)(10.369,7.000){2}{\rule{0.393pt}{0.400pt}}
\multiput(681.00,402.59)(0.692,0.488){13}{\rule{0.650pt}{0.117pt}}
\multiput(681.00,401.17)(9.651,8.000){2}{\rule{0.325pt}{0.400pt}}
\multiput(692.00,410.59)(0.874,0.485){11}{\rule{0.786pt}{0.117pt}}
\multiput(692.00,409.17)(10.369,7.000){2}{\rule{0.393pt}{0.400pt}}
\multiput(704.00,417.59)(0.758,0.488){13}{\rule{0.700pt}{0.117pt}}
\multiput(704.00,416.17)(10.547,8.000){2}{\rule{0.350pt}{0.400pt}}
\multiput(716.00,425.59)(0.874,0.485){11}{\rule{0.786pt}{0.117pt}}
\multiput(716.00,424.17)(10.369,7.000){2}{\rule{0.393pt}{0.400pt}}
\multiput(728.00,432.59)(0.874,0.485){11}{\rule{0.786pt}{0.117pt}}
\multiput(728.00,431.17)(10.369,7.000){2}{\rule{0.393pt}{0.400pt}}
\multiput(740.00,439.59)(0.692,0.488){13}{\rule{0.650pt}{0.117pt}}
\multiput(740.00,438.17)(9.651,8.000){2}{\rule{0.325pt}{0.400pt}}
\multiput(751.00,447.59)(0.874,0.485){11}{\rule{0.786pt}{0.117pt}}
\multiput(751.00,446.17)(10.369,7.000){2}{\rule{0.393pt}{0.400pt}}
\multiput(763.00,454.59)(0.758,0.488){13}{\rule{0.700pt}{0.117pt}}
\multiput(763.00,453.17)(10.547,8.000){2}{\rule{0.350pt}{0.400pt}}
\multiput(775.00,462.59)(0.874,0.485){11}{\rule{0.786pt}{0.117pt}}
\multiput(775.00,461.17)(10.369,7.000){2}{\rule{0.393pt}{0.400pt}}
\multiput(787.00,469.59)(0.874,0.485){11}{\rule{0.786pt}{0.117pt}}
\multiput(787.00,468.17)(10.369,7.000){2}{\rule{0.393pt}{0.400pt}}
\multiput(799.00,476.59)(0.692,0.488){13}{\rule{0.650pt}{0.117pt}}
\multiput(799.00,475.17)(9.651,8.000){2}{\rule{0.325pt}{0.400pt}}
\multiput(810.00,484.59)(0.874,0.485){11}{\rule{0.786pt}{0.117pt}}
\multiput(810.00,483.17)(10.369,7.000){2}{\rule{0.393pt}{0.400pt}}
\multiput(822.00,491.59)(0.758,0.488){13}{\rule{0.700pt}{0.117pt}}
\multiput(822.00,490.17)(10.547,8.000){2}{\rule{0.350pt}{0.400pt}}
\multiput(834.00,499.59)(0.874,0.485){11}{\rule{0.786pt}{0.117pt}}
\multiput(834.00,498.17)(10.369,7.000){2}{\rule{0.393pt}{0.400pt}}
\multiput(846.00,506.59)(0.692,0.488){13}{\rule{0.650pt}{0.117pt}}
\multiput(846.00,505.17)(9.651,8.000){2}{\rule{0.325pt}{0.400pt}}
\multiput(857.00,514.59)(0.874,0.485){11}{\rule{0.786pt}{0.117pt}}
\multiput(857.00,513.17)(10.369,7.000){2}{\rule{0.393pt}{0.400pt}}
\multiput(869.00,521.59)(0.874,0.485){11}{\rule{0.786pt}{0.117pt}}
\multiput(869.00,520.17)(10.369,7.000){2}{\rule{0.393pt}{0.400pt}}
\multiput(881.00,528.59)(0.758,0.488){13}{\rule{0.700pt}{0.117pt}}
\multiput(881.00,527.17)(10.547,8.000){2}{\rule{0.350pt}{0.400pt}}
\multiput(893.00,536.59)(0.874,0.485){11}{\rule{0.786pt}{0.117pt}}
\multiput(893.00,535.17)(10.369,7.000){2}{\rule{0.393pt}{0.400pt}}
\multiput(905.00,543.59)(0.692,0.488){13}{\rule{0.650pt}{0.117pt}}
\multiput(905.00,542.17)(9.651,8.000){2}{\rule{0.325pt}{0.400pt}}
\multiput(916.00,551.59)(0.874,0.485){11}{\rule{0.786pt}{0.117pt}}
\multiput(916.00,550.17)(10.369,7.000){2}{\rule{0.393pt}{0.400pt}}
\multiput(928.00,558.59)(0.874,0.485){11}{\rule{0.786pt}{0.117pt}}
\multiput(928.00,557.17)(10.369,7.000){2}{\rule{0.393pt}{0.400pt}}
\multiput(940.00,565.59)(0.758,0.488){13}{\rule{0.700pt}{0.117pt}}
\multiput(940.00,564.17)(10.547,8.000){2}{\rule{0.350pt}{0.400pt}}
\multiput(952.00,573.59)(0.874,0.485){11}{\rule{0.786pt}{0.117pt}}
\multiput(952.00,572.17)(10.369,7.000){2}{\rule{0.393pt}{0.400pt}}
\multiput(964.00,580.59)(0.692,0.488){13}{\rule{0.650pt}{0.117pt}}
\multiput(964.00,579.17)(9.651,8.000){2}{\rule{0.325pt}{0.400pt}}
\multiput(975.00,588.59)(0.874,0.485){11}{\rule{0.786pt}{0.117pt}}
\multiput(975.00,587.17)(10.369,7.000){2}{\rule{0.393pt}{0.400pt}}
\multiput(987.00,595.59)(0.874,0.485){11}{\rule{0.786pt}{0.117pt}}
\multiput(987.00,594.17)(10.369,7.000){2}{\rule{0.393pt}{0.400pt}}
\multiput(999.00,602.59)(0.758,0.488){13}{\rule{0.700pt}{0.117pt}}
\multiput(999.00,601.17)(10.547,8.000){2}{\rule{0.350pt}{0.400pt}}
\multiput(1011.00,610.59)(0.874,0.485){11}{\rule{0.786pt}{0.117pt}}
\multiput(1011.00,609.17)(10.369,7.000){2}{\rule{0.393pt}{0.400pt}}
\multiput(1023.00,617.59)(0.692,0.488){13}{\rule{0.650pt}{0.117pt}}
\multiput(1023.00,616.17)(9.651,8.000){2}{\rule{0.325pt}{0.400pt}}
\multiput(1034.00,625.59)(0.874,0.485){11}{\rule{0.786pt}{0.117pt}}
\multiput(1034.00,624.17)(10.369,7.000){2}{\rule{0.393pt}{0.400pt}}
\multiput(1046.00,632.59)(0.874,0.485){11}{\rule{0.786pt}{0.117pt}}
\multiput(1046.00,631.17)(10.369,7.000){2}{\rule{0.393pt}{0.400pt}}
\multiput(1058.00,639.59)(0.758,0.488){13}{\rule{0.700pt}{0.117pt}}
\multiput(1058.00,638.17)(10.547,8.000){2}{\rule{0.350pt}{0.400pt}}
\multiput(1070.00,647.59)(0.874,0.485){11}{\rule{0.786pt}{0.117pt}}
\multiput(1070.00,646.17)(10.369,7.000){2}{\rule{0.393pt}{0.400pt}}
\multiput(1082.00,654.59)(0.692,0.488){13}{\rule{0.650pt}{0.117pt}}
\multiput(1082.00,653.17)(9.651,8.000){2}{\rule{0.325pt}{0.400pt}}
\multiput(1093.00,662.59)(0.874,0.485){11}{\rule{0.786pt}{0.117pt}}
\multiput(1093.00,661.17)(10.369,7.000){2}{\rule{0.393pt}{0.400pt}}
\multiput(1105.00,669.59)(0.758,0.488){13}{\rule{0.700pt}{0.117pt}}
\multiput(1105.00,668.17)(10.547,8.000){2}{\rule{0.350pt}{0.400pt}}
\multiput(1117.00,677.59)(0.874,0.485){11}{\rule{0.786pt}{0.117pt}}
\multiput(1117.00,676.17)(10.369,7.000){2}{\rule{0.393pt}{0.400pt}}
\multiput(1129.00,684.59)(0.798,0.485){11}{\rule{0.729pt}{0.117pt}}
\multiput(1129.00,683.17)(9.488,7.000){2}{\rule{0.364pt}{0.400pt}}
\multiput(1140.00,691.59)(0.758,0.488){13}{\rule{0.700pt}{0.117pt}}
\multiput(1140.00,690.17)(10.547,8.000){2}{\rule{0.350pt}{0.400pt}}
\multiput(1152.00,699.59)(0.874,0.485){11}{\rule{0.786pt}{0.117pt}}
\multiput(1152.00,698.17)(10.369,7.000){2}{\rule{0.393pt}{0.400pt}}
\multiput(1164.00,706.59)(0.758,0.488){13}{\rule{0.700pt}{0.117pt}}
\multiput(1164.00,705.17)(10.547,8.000){2}{\rule{0.350pt}{0.400pt}}
\multiput(1176.00,714.59)(0.874,0.485){11}{\rule{0.786pt}{0.117pt}}
\multiput(1176.00,713.17)(10.369,7.000){2}{\rule{0.393pt}{0.400pt}}
\multiput(1188.00,721.59)(0.798,0.485){11}{\rule{0.729pt}{0.117pt}}
\multiput(1188.00,720.17)(9.488,7.000){2}{\rule{0.364pt}{0.400pt}}
\multiput(1199.00,728.59)(0.758,0.488){13}{\rule{0.700pt}{0.117pt}}
\multiput(1199.00,727.17)(10.547,8.000){2}{\rule{0.350pt}{0.400pt}}
\multiput(1211.00,736.59)(0.874,0.485){11}{\rule{0.786pt}{0.117pt}}
\multiput(1211.00,735.17)(10.369,7.000){2}{\rule{0.393pt}{0.400pt}}
\multiput(1223.00,743.59)(0.758,0.488){13}{\rule{0.700pt}{0.117pt}}
\multiput(1223.00,742.17)(10.547,8.000){2}{\rule{0.350pt}{0.400pt}}
\multiput(1235.00,751.59)(0.874,0.485){11}{\rule{0.786pt}{0.117pt}}
\multiput(1235.00,750.17)(10.369,7.000){2}{\rule{0.393pt}{0.400pt}}
\multiput(1247.00,758.59)(0.798,0.485){11}{\rule{0.729pt}{0.117pt}}
\multiput(1247.00,757.17)(9.488,7.000){2}{\rule{0.364pt}{0.400pt}}
\multiput(1258.00,765.59)(0.758,0.488){13}{\rule{0.700pt}{0.117pt}}
\multiput(1258.00,764.17)(10.547,8.000){2}{\rule{0.350pt}{0.400pt}}
\multiput(1270.00,773.59)(0.874,0.485){11}{\rule{0.786pt}{0.117pt}}
\multiput(1270.00,772.17)(10.369,7.000){2}{\rule{0.393pt}{0.400pt}}
\multiput(1282.00,780.59)(0.758,0.488){13}{\rule{0.700pt}{0.117pt}}
\multiput(1282.00,779.17)(10.547,8.000){2}{\rule{0.350pt}{0.400pt}}
\multiput(1294.00,788.59)(0.874,0.485){11}{\rule{0.786pt}{0.117pt}}
\multiput(1294.00,787.17)(10.369,7.000){2}{\rule{0.393pt}{0.400pt}}
\multiput(1306.00,795.59)(0.798,0.485){11}{\rule{0.729pt}{0.117pt}}
\multiput(1306.00,794.17)(9.488,7.000){2}{\rule{0.364pt}{0.400pt}}
\multiput(1317.00,802.59)(0.758,0.488){13}{\rule{0.700pt}{0.117pt}}
\multiput(1317.00,801.17)(10.547,8.000){2}{\rule{0.350pt}{0.400pt}}
\multiput(1329.00,810.59)(0.874,0.485){11}{\rule{0.786pt}{0.117pt}}
\multiput(1329.00,809.17)(10.369,7.000){2}{\rule{0.393pt}{0.400pt}}
\multiput(1341.00,817.59)(0.758,0.488){13}{\rule{0.700pt}{0.117pt}}
\multiput(1341.00,816.17)(10.547,8.000){2}{\rule{0.350pt}{0.400pt}}
\multiput(1353.00,825.59)(0.874,0.485){11}{\rule{0.786pt}{0.117pt}}
\multiput(1353.00,824.17)(10.369,7.000){2}{\rule{0.393pt}{0.400pt}}
\multiput(1365.00,832.59)(0.798,0.485){11}{\rule{0.729pt}{0.117pt}}
\multiput(1365.00,831.17)(9.488,7.000){2}{\rule{0.364pt}{0.400pt}}
\multiput(1376.00,839.59)(0.758,0.488){13}{\rule{0.700pt}{0.117pt}}
\multiput(1376.00,838.17)(10.547,8.000){2}{\rule{0.350pt}{0.400pt}}
\multiput(1388.00,847.59)(0.874,0.485){11}{\rule{0.786pt}{0.117pt}}
\multiput(1388.00,846.17)(10.369,7.000){2}{\rule{0.393pt}{0.400pt}}
\multiput(1400.00,854.59)(0.758,0.488){13}{\rule{0.700pt}{0.117pt}}
\multiput(1400.00,853.17)(10.547,8.000){2}{\rule{0.350pt}{0.400pt}}
\end{picture}

\caption{Average temperature of the particles in each of the 25 bins
in which the $x$ axis of the box has been divided. Solid line is the
theoretical Fourier profile. Open circles correspond to $N=100$ and
solid circles correspond to $N=1000$ particles within the box.
Other parameters are $s=1.5$, $C_L=10^8$.
}
\label{figgrad}
\end{center}
\end{figure}
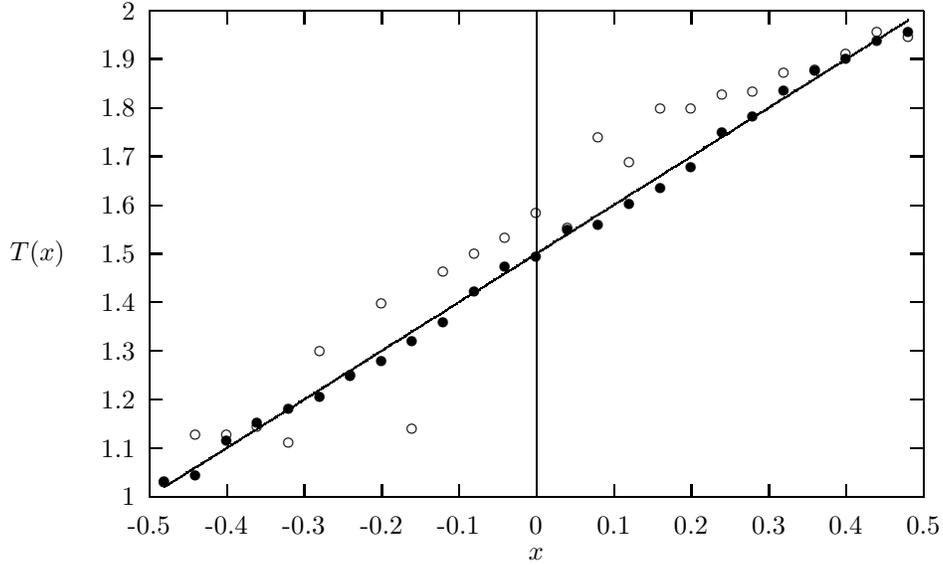

\begin{figure}[ht]
\begin{center}
\setlength{\unitlength}{0.240900pt}
\ifx\plotpoint\undefined\newsavebox{\plotpoint}\fi
\begin{picture}(1500,900)(0,0)
\font\gnuplot=cmr10 at 10pt
\gnuplot
\sbox{\plotpoint}{\rule[-0.200pt]{0.400pt}{0.400pt}}%
\put(220.0,113.0){\rule[-0.200pt]{292.934pt}{0.400pt}}
\put(220.0,113.0){\rule[-0.200pt]{0.400pt}{184.048pt}}
\put(220.0,113.0){\rule[-0.200pt]{4.818pt}{0.400pt}}
\put(198,113){\makebox(0,0)[r]{0}}
\put(1416.0,113.0){\rule[-0.200pt]{4.818pt}{0.400pt}}
\put(220.0,240.0){\rule[-0.200pt]{4.818pt}{0.400pt}}
\put(198,240){\makebox(0,0)[r]{0.2}}
\put(1416.0,240.0){\rule[-0.200pt]{4.818pt}{0.400pt}}
\put(220.0,368.0){\rule[-0.200pt]{4.818pt}{0.400pt}}
\put(198,368){\makebox(0,0)[r]{0.4}}
\put(1416.0,368.0){\rule[-0.200pt]{4.818pt}{0.400pt}}
\put(220.0,495.0){\rule[-0.200pt]{4.818pt}{0.400pt}}
\put(198,495){\makebox(0,0)[r]{0.6}}
\put(1416.0,495.0){\rule[-0.200pt]{4.818pt}{0.400pt}}
\put(220.0,622.0){\rule[-0.200pt]{4.818pt}{0.400pt}}
\put(198,622){\makebox(0,0)[r]{0.8}}
\put(1416.0,622.0){\rule[-0.200pt]{4.818pt}{0.400pt}}
\put(220.0,750.0){\rule[-0.200pt]{4.818pt}{0.400pt}}
\put(198,750){\makebox(0,0)[r]{1}}
\put(1416.0,750.0){\rule[-0.200pt]{4.818pt}{0.400pt}}
\put(220.0,877.0){\rule[-0.200pt]{4.818pt}{0.400pt}}
\put(198,877){\makebox(0,0)[r]{1.2}}
\put(1416.0,877.0){\rule[-0.200pt]{4.818pt}{0.400pt}}
\put(220.0,113.0){\rule[-0.200pt]{0.400pt}{4.818pt}}
\put(220,68){\makebox(0,0){0}}
\put(220.0,857.0){\rule[-0.200pt]{0.400pt}{4.818pt}}
\put(510.0,113.0){\rule[-0.200pt]{0.400pt}{4.818pt}}
\put(510,68){\makebox(0,0){5}}
\put(510.0,857.0){\rule[-0.200pt]{0.400pt}{4.818pt}}
\put(799.0,113.0){\rule[-0.200pt]{0.400pt}{4.818pt}}
\put(799,68){\makebox(0,0){10}}
\put(799.0,857.0){\rule[-0.200pt]{0.400pt}{4.818pt}}
\put(1089.0,113.0){\rule[-0.200pt]{0.400pt}{4.818pt}}
\put(1089,68){\makebox(0,0){15}}
\put(1089.0,857.0){\rule[-0.200pt]{0.400pt}{4.818pt}}
\put(1378.0,113.0){\rule[-0.200pt]{0.400pt}{4.818pt}}
\put(1378,68){\makebox(0,0){20}}
\put(1378.0,857.0){\rule[-0.200pt]{0.400pt}{4.818pt}}
\put(220.0,113.0){\rule[-0.200pt]{292.934pt}{0.400pt}}
\put(1436.0,113.0){\rule[-0.200pt]{0.400pt}{184.048pt}}
\put(220.0,877.0){\rule[-0.200pt]{292.934pt}{0.400pt}}
\put(45,495){\makebox(0,0){${\bf Q}_x$}}
\put(828,23){\makebox(0,0){$(\nabla T)_x$}}
\put(220.0,113.0){\rule[-0.200pt]{0.400pt}{184.048pt}}
\sbox{\plotpoint}{\rule[-0.400pt]{0.800pt}{0.800pt}}%
\put(278,144){\raisebox{-.8pt}{\makebox(0,0){$\Diamond$}}}
\put(336,177){\raisebox{-.8pt}{\makebox(0,0){$\Diamond$}}}
\put(394,209){\raisebox{-.8pt}{\makebox(0,0){$\Diamond$}}}
\put(452,241){\raisebox{-.8pt}{\makebox(0,0){$\Diamond$}}}
\put(510,271){\raisebox{-.8pt}{\makebox(0,0){$\Diamond$}}}
\put(567,302){\raisebox{-.8pt}{\makebox(0,0){$\Diamond$}}}
\put(683,366){\raisebox{-.8pt}{\makebox(0,0){$\Diamond$}}}
\put(799,432){\raisebox{-.8pt}{\makebox(0,0){$\Diamond$}}}
\put(1089,598){\raisebox{-.8pt}{\makebox(0,0){$\Diamond$}}}
\put(1378,763){\raisebox{-.8pt}{\makebox(0,0){$\Diamond$}}}
\sbox{\plotpoint}{\rule[-0.200pt]{0.400pt}{0.400pt}}%
\put(220,113){\usebox{\plotpoint}}
\put(220.00,113.00){\usebox{\plotpoint}}
\put(238.23,122.88){\usebox{\plotpoint}}
\put(256.49,132.70){\usebox{\plotpoint}}
\multiput(257,133)(17.928,10.458){0}{\usebox{\plotpoint}}
\put(274.61,142.80){\usebox{\plotpoint}}
\put(292.98,152.45){\usebox{\plotpoint}}
\multiput(294,153)(17.928,10.458){0}{\usebox{\plotpoint}}
\put(311.11,162.55){\usebox{\plotpoint}}
\put(329.49,172.19){\usebox{\plotpoint}}
\multiput(331,173)(17.928,10.458){0}{\usebox{\plotpoint}}
\put(347.60,182.30){\usebox{\plotpoint}}
\put(365.78,192.29){\usebox{\plotpoint}}
\multiput(367,193)(18.275,9.840){0}{\usebox{\plotpoint}}
\put(384.10,202.05){\usebox{\plotpoint}}
\put(402.30,212.01){\usebox{\plotpoint}}
\multiput(404,213)(18.275,9.840){0}{\usebox{\plotpoint}}
\put(420.60,221.80){\usebox{\plotpoint}}
\put(438.81,231.72){\usebox{\plotpoint}}
\multiput(441,233)(17.928,10.458){0}{\usebox{\plotpoint}}
\put(456.93,241.81){\usebox{\plotpoint}}
\put(475.30,251.43){\usebox{\plotpoint}}
\multiput(478,253)(17.928,10.458){0}{\usebox{\plotpoint}}
\put(493.39,261.57){\usebox{\plotpoint}}
\put(511.79,271.13){\usebox{\plotpoint}}
\multiput(515,273)(17.928,10.458){0}{\usebox{\plotpoint}}
\put(529.81,281.41){\usebox{\plotpoint}}
\put(548.23,290.97){\usebox{\plotpoint}}
\multiput(552,293)(17.928,10.458){0}{\usebox{\plotpoint}}
\put(566.31,301.16){\usebox{\plotpoint}}
\put(584.57,311.00){\usebox{\plotpoint}}
\multiput(588,313)(18.275,9.840){0}{\usebox{\plotpoint}}
\put(602.81,320.90){\usebox{\plotpoint}}
\put(621.09,330.72){\usebox{\plotpoint}}
\multiput(625,333)(18.275,9.840){0}{\usebox{\plotpoint}}
\put(639.30,340.65){\usebox{\plotpoint}}
\put(657.60,350.43){\usebox{\plotpoint}}
\multiput(662,353)(17.928,10.458){0}{\usebox{\plotpoint}}
\put(675.61,360.74){\usebox{\plotpoint}}
\put(694.09,370.13){\usebox{\plotpoint}}
\multiput(699,373)(17.928,10.458){0}{\usebox{\plotpoint}}
\put(712.04,380.56){\usebox{\plotpoint}}
\put(730.41,390.21){\usebox{\plotpoint}}
\multiput(736,393)(17.928,10.458){0}{\usebox{\plotpoint}}
\put(748.53,400.31){\usebox{\plotpoint}}
\put(766.79,410.13){\usebox{\plotpoint}}
\multiput(773,413)(17.928,10.458){0}{\usebox{\plotpoint}}
\put(785.02,420.01){\usebox{\plotpoint}}
\put(803.25,429.89){\usebox{\plotpoint}}
\put(821.51,439.71){\usebox{\plotpoint}}
\multiput(822,440)(17.928,10.458){0}{\usebox{\plotpoint}}
\put(839.63,449.81){\usebox{\plotpoint}}
\put(858.00,459.46){\usebox{\plotpoint}}
\multiput(859,460)(17.928,10.458){0}{\usebox{\plotpoint}}
\put(876.13,469.56){\usebox{\plotpoint}}
\put(894.51,479.20){\usebox{\plotpoint}}
\multiput(896,480)(17.928,10.458){0}{\usebox{\plotpoint}}
\put(912.62,489.31){\usebox{\plotpoint}}
\put(930.80,499.30){\usebox{\plotpoint}}
\multiput(932,500)(18.275,9.840){0}{\usebox{\plotpoint}}
\put(949.12,509.06){\usebox{\plotpoint}}
\put(967.32,519.02){\usebox{\plotpoint}}
\multiput(969,520)(18.275,9.840){0}{\usebox{\plotpoint}}
\put(985.62,528.81){\usebox{\plotpoint}}
\put(1003.83,538.74){\usebox{\plotpoint}}
\multiput(1006,540)(17.928,10.458){0}{\usebox{\plotpoint}}
\put(1021.95,548.82){\usebox{\plotpoint}}
\put(1040.32,558.44){\usebox{\plotpoint}}
\multiput(1043,560)(17.928,10.458){0}{\usebox{\plotpoint}}
\put(1058.42,568.58){\usebox{\plotpoint}}
\put(1076.81,578.14){\usebox{\plotpoint}}
\multiput(1080,580)(17.928,10.458){0}{\usebox{\plotpoint}}
\put(1094.83,588.42){\usebox{\plotpoint}}
\put(1113.25,597.98){\usebox{\plotpoint}}
\multiput(1117,600)(17.928,10.458){0}{\usebox{\plotpoint}}
\put(1131.33,608.17){\usebox{\plotpoint}}
\put(1149.59,618.01){\usebox{\plotpoint}}
\multiput(1153,620)(18.275,9.840){0}{\usebox{\plotpoint}}
\put(1167.83,627.91){\usebox{\plotpoint}}
\put(1186.11,637.73){\usebox{\plotpoint}}
\multiput(1190,640)(18.275,9.840){0}{\usebox{\plotpoint}}
\put(1204.33,647.66){\usebox{\plotpoint}}
\put(1222.62,657.44){\usebox{\plotpoint}}
\multiput(1227,660)(17.928,10.458){0}{\usebox{\plotpoint}}
\put(1240.63,667.75){\usebox{\plotpoint}}
\put(1259.11,677.15){\usebox{\plotpoint}}
\multiput(1264,680)(17.928,10.458){0}{\usebox{\plotpoint}}
\put(1277.09,687.50){\usebox{\plotpoint}}
\put(1295.60,696.85){\usebox{\plotpoint}}
\multiput(1301,700)(17.928,10.458){0}{\usebox{\plotpoint}}
\put(1313.54,707.27){\usebox{\plotpoint}}
\put(1332.00,716.77){\usebox{\plotpoint}}
\multiput(1338,720)(17.928,10.458){0}{\usebox{\plotpoint}}
\put(1350.04,727.02){\usebox{\plotpoint}}
\put(1368.50,736.50){\usebox{\plotpoint}}
\put(1386.55,746.74){\usebox{\plotpoint}}
\multiput(1387,747)(18.564,9.282){0}{\usebox{\plotpoint}}
\put(1404.89,756.44){\usebox{\plotpoint}}
\put(1423.05,766.49){\usebox{\plotpoint}}
\multiput(1424,767)(18.564,9.282){0}{\usebox{\plotpoint}}
\put(1436,773){\usebox{\plotpoint}}
\end{picture}

\caption{Average $x$ component of the heat flux as a function
of the imposed temperature gradient. A linear dependence is obtained
with small deviations at large gradients. The slope of the straight
line gives the thermal diffusivity.
}
\label{figq}
\end{center}
\end{figure}
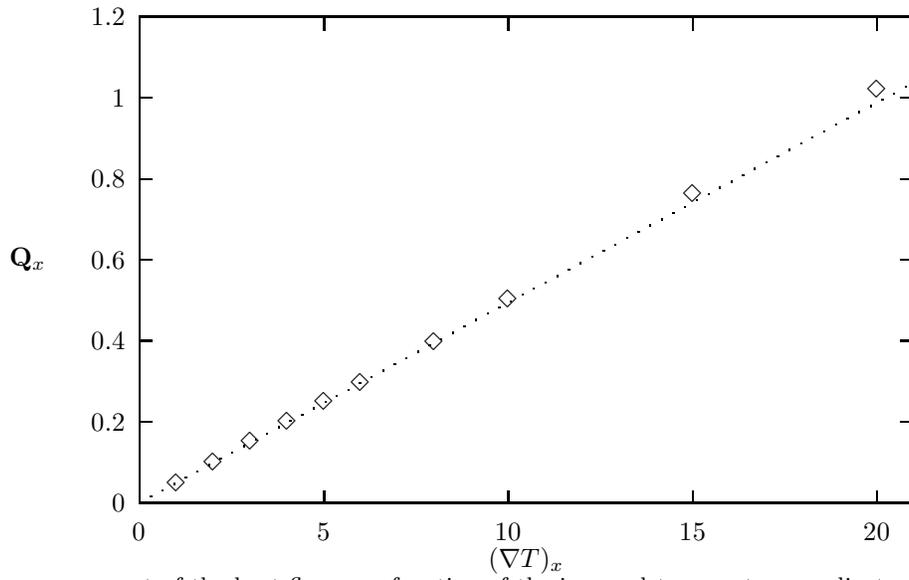

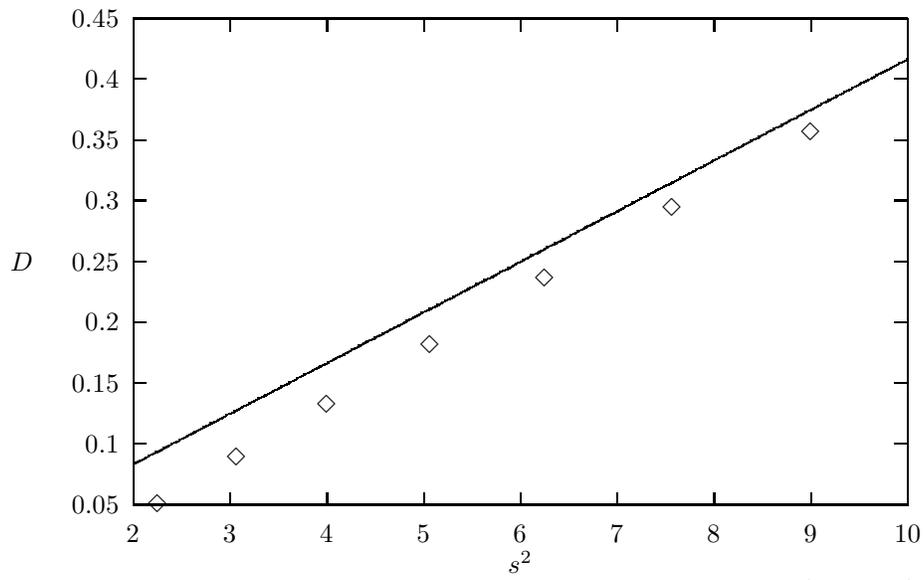
\begin{figure}[ht]
\begin{center}
\setlength{\unitlength}{0.240900pt}
\ifx\plotpoint\undefined\newsavebox{\plotpoint}\fi
\begin{picture}(1500,900)(0,0)
\font\gnuplot=cmr10 at 10pt
\gnuplot
\sbox{\plotpoint}{\rule[-0.200pt]{0.400pt}{0.400pt}}%
\put(220.0,113.0){\rule[-0.200pt]{4.818pt}{0.400pt}}
\put(198,113){\makebox(0,0)[r]{0.05}}
\put(1416.0,113.0){\rule[-0.200pt]{4.818pt}{0.400pt}}
\put(220.0,209.0){\rule[-0.200pt]{4.818pt}{0.400pt}}
\put(198,209){\makebox(0,0)[r]{0.1}}
\put(1416.0,209.0){\rule[-0.200pt]{4.818pt}{0.400pt}}
\put(220.0,304.0){\rule[-0.200pt]{4.818pt}{0.400pt}}
\put(198,304){\makebox(0,0)[r]{0.15}}
\put(1416.0,304.0){\rule[-0.200pt]{4.818pt}{0.400pt}}
\put(220.0,400.0){\rule[-0.200pt]{4.818pt}{0.400pt}}
\put(198,400){\makebox(0,0)[r]{0.2}}
\put(1416.0,400.0){\rule[-0.200pt]{4.818pt}{0.400pt}}
\put(220.0,495.0){\rule[-0.200pt]{4.818pt}{0.400pt}}
\put(198,495){\makebox(0,0)[r]{0.25}}
\put(1416.0,495.0){\rule[-0.200pt]{4.818pt}{0.400pt}}
\put(220.0,591.0){\rule[-0.200pt]{4.818pt}{0.400pt}}
\put(198,591){\makebox(0,0)[r]{0.3}}
\put(1416.0,591.0){\rule[-0.200pt]{4.818pt}{0.400pt}}
\put(220.0,686.0){\rule[-0.200pt]{4.818pt}{0.400pt}}
\put(198,686){\makebox(0,0)[r]{0.35}}
\put(1416.0,686.0){\rule[-0.200pt]{4.818pt}{0.400pt}}
\put(220.0,782.0){\rule[-0.200pt]{4.818pt}{0.400pt}}
\put(198,782){\makebox(0,0)[r]{0.4}}
\put(1416.0,782.0){\rule[-0.200pt]{4.818pt}{0.400pt}}
\put(220.0,877.0){\rule[-0.200pt]{4.818pt}{0.400pt}}
\put(198,877){\makebox(0,0)[r]{0.45}}
\put(1416.0,877.0){\rule[-0.200pt]{4.818pt}{0.400pt}}
\put(220.0,113.0){\rule[-0.200pt]{0.400pt}{4.818pt}}
\put(220,68){\makebox(0,0){2}}
\put(220.0,857.0){\rule[-0.200pt]{0.400pt}{4.818pt}}
\put(372.0,113.0){\rule[-0.200pt]{0.400pt}{4.818pt}}
\put(372,68){\makebox(0,0){3}}
\put(372.0,857.0){\rule[-0.200pt]{0.400pt}{4.818pt}}
\put(524.0,113.0){\rule[-0.200pt]{0.400pt}{4.818pt}}
\put(524,68){\makebox(0,0){4}}
\put(524.0,857.0){\rule[-0.200pt]{0.400pt}{4.818pt}}
\put(676.0,113.0){\rule[-0.200pt]{0.400pt}{4.818pt}}
\put(676,68){\makebox(0,0){5}}
\put(676.0,857.0){\rule[-0.200pt]{0.400pt}{4.818pt}}
\put(828.0,113.0){\rule[-0.200pt]{0.400pt}{4.818pt}}
\put(828,68){\makebox(0,0){6}}
\put(828.0,857.0){\rule[-0.200pt]{0.400pt}{4.818pt}}
\put(980.0,113.0){\rule[-0.200pt]{0.400pt}{4.818pt}}
\put(980,68){\makebox(0,0){7}}
\put(980.0,857.0){\rule[-0.200pt]{0.400pt}{4.818pt}}
\put(1132.0,113.0){\rule[-0.200pt]{0.400pt}{4.818pt}}
\put(1132,68){\makebox(0,0){8}}
\put(1132.0,857.0){\rule[-0.200pt]{0.400pt}{4.818pt}}
\put(1284.0,113.0){\rule[-0.200pt]{0.400pt}{4.818pt}}
\put(1284,68){\makebox(0,0){9}}
\put(1284.0,857.0){\rule[-0.200pt]{0.400pt}{4.818pt}}
\put(1436.0,113.0){\rule[-0.200pt]{0.400pt}{4.818pt}}
\put(1436,68){\makebox(0,0){10}}
\put(1436.0,857.0){\rule[-0.200pt]{0.400pt}{4.818pt}}
\put(220.0,113.0){\rule[-0.200pt]{292.934pt}{0.400pt}}
\put(1436.0,113.0){\rule[-0.200pt]{0.400pt}{184.048pt}}
\put(220.0,877.0){\rule[-0.200pt]{292.934pt}{0.400pt}}
\put(45,495){\makebox(0,0){$ D$}}
\put(828,23){\makebox(0,0){$s^2$}}
\put(220.0,113.0){\rule[-0.200pt]{0.400pt}{184.048pt}}
\put(258,114){\raisebox{-.8pt}{\makebox(0,0){$\Diamond$}}}
\put(382,188){\raisebox{-.8pt}{\makebox(0,0){$\Diamond$}}}
\put(524,271){\raisebox{-.8pt}{\makebox(0,0){$\Diamond$}}}
\put(686,365){\raisebox{-.8pt}{\makebox(0,0){$\Diamond$}}}
\put(866,469){\raisebox{-.8pt}{\makebox(0,0){$\Diamond$}}}
\put(1066,580){\raisebox{-.8pt}{\makebox(0,0){$\Diamond$}}}
\put(1284,699){\raisebox{-.8pt}{\makebox(0,0){$\Diamond$}}}
\put(220,177){\usebox{\plotpoint}}
\multiput(220.00,177.59)(1.033,0.482){9}{\rule{0.900pt}{0.116pt}}
\multiput(220.00,176.17)(10.132,6.000){2}{\rule{0.450pt}{0.400pt}}
\multiput(232.00,183.59)(0.950,0.485){11}{\rule{0.843pt}{0.117pt}}
\multiput(232.00,182.17)(11.251,7.000){2}{\rule{0.421pt}{0.400pt}}
\multiput(245.00,190.59)(1.033,0.482){9}{\rule{0.900pt}{0.116pt}}
\multiput(245.00,189.17)(10.132,6.000){2}{\rule{0.450pt}{0.400pt}}
\multiput(257.00,196.59)(1.033,0.482){9}{\rule{0.900pt}{0.116pt}}
\multiput(257.00,195.17)(10.132,6.000){2}{\rule{0.450pt}{0.400pt}}
\multiput(269.00,202.59)(0.874,0.485){11}{\rule{0.786pt}{0.117pt}}
\multiput(269.00,201.17)(10.369,7.000){2}{\rule{0.393pt}{0.400pt}}
\multiput(281.00,209.59)(1.123,0.482){9}{\rule{0.967pt}{0.116pt}}
\multiput(281.00,208.17)(10.994,6.000){2}{\rule{0.483pt}{0.400pt}}
\multiput(294.00,215.59)(0.874,0.485){11}{\rule{0.786pt}{0.117pt}}
\multiput(294.00,214.17)(10.369,7.000){2}{\rule{0.393pt}{0.400pt}}
\multiput(306.00,222.59)(1.033,0.482){9}{\rule{0.900pt}{0.116pt}}
\multiput(306.00,221.17)(10.132,6.000){2}{\rule{0.450pt}{0.400pt}}
\multiput(318.00,228.59)(0.950,0.485){11}{\rule{0.843pt}{0.117pt}}
\multiput(318.00,227.17)(11.251,7.000){2}{\rule{0.421pt}{0.400pt}}
\multiput(331.00,235.59)(1.033,0.482){9}{\rule{0.900pt}{0.116pt}}
\multiput(331.00,234.17)(10.132,6.000){2}{\rule{0.450pt}{0.400pt}}
\multiput(343.00,241.59)(1.033,0.482){9}{\rule{0.900pt}{0.116pt}}
\multiput(343.00,240.17)(10.132,6.000){2}{\rule{0.450pt}{0.400pt}}
\multiput(355.00,247.59)(0.874,0.485){11}{\rule{0.786pt}{0.117pt}}
\multiput(355.00,246.17)(10.369,7.000){2}{\rule{0.393pt}{0.400pt}}
\multiput(367.00,254.59)(1.123,0.482){9}{\rule{0.967pt}{0.116pt}}
\multiput(367.00,253.17)(10.994,6.000){2}{\rule{0.483pt}{0.400pt}}
\multiput(380.00,260.59)(0.874,0.485){11}{\rule{0.786pt}{0.117pt}}
\multiput(380.00,259.17)(10.369,7.000){2}{\rule{0.393pt}{0.400pt}}
\multiput(392.00,267.59)(1.033,0.482){9}{\rule{0.900pt}{0.116pt}}
\multiput(392.00,266.17)(10.132,6.000){2}{\rule{0.450pt}{0.400pt}}
\multiput(404.00,273.59)(0.950,0.485){11}{\rule{0.843pt}{0.117pt}}
\multiput(404.00,272.17)(11.251,7.000){2}{\rule{0.421pt}{0.400pt}}
\multiput(417.00,280.59)(1.033,0.482){9}{\rule{0.900pt}{0.116pt}}
\multiput(417.00,279.17)(10.132,6.000){2}{\rule{0.450pt}{0.400pt}}
\multiput(429.00,286.59)(1.033,0.482){9}{\rule{0.900pt}{0.116pt}}
\multiput(429.00,285.17)(10.132,6.000){2}{\rule{0.450pt}{0.400pt}}
\multiput(441.00,292.59)(0.874,0.485){11}{\rule{0.786pt}{0.117pt}}
\multiput(441.00,291.17)(10.369,7.000){2}{\rule{0.393pt}{0.400pt}}
\multiput(453.00,299.59)(1.123,0.482){9}{\rule{0.967pt}{0.116pt}}
\multiput(453.00,298.17)(10.994,6.000){2}{\rule{0.483pt}{0.400pt}}
\multiput(466.00,305.59)(0.874,0.485){11}{\rule{0.786pt}{0.117pt}}
\multiput(466.00,304.17)(10.369,7.000){2}{\rule{0.393pt}{0.400pt}}
\multiput(478.00,312.59)(1.033,0.482){9}{\rule{0.900pt}{0.116pt}}
\multiput(478.00,311.17)(10.132,6.000){2}{\rule{0.450pt}{0.400pt}}
\multiput(490.00,318.59)(0.950,0.485){11}{\rule{0.843pt}{0.117pt}}
\multiput(490.00,317.17)(11.251,7.000){2}{\rule{0.421pt}{0.400pt}}
\multiput(503.00,325.59)(1.033,0.482){9}{\rule{0.900pt}{0.116pt}}
\multiput(503.00,324.17)(10.132,6.000){2}{\rule{0.450pt}{0.400pt}}
\multiput(515.00,331.59)(1.033,0.482){9}{\rule{0.900pt}{0.116pt}}
\multiput(515.00,330.17)(10.132,6.000){2}{\rule{0.450pt}{0.400pt}}
\multiput(527.00,337.59)(0.874,0.485){11}{\rule{0.786pt}{0.117pt}}
\multiput(527.00,336.17)(10.369,7.000){2}{\rule{0.393pt}{0.400pt}}
\multiput(539.00,344.59)(1.123,0.482){9}{\rule{0.967pt}{0.116pt}}
\multiput(539.00,343.17)(10.994,6.000){2}{\rule{0.483pt}{0.400pt}}
\multiput(552.00,350.59)(0.874,0.485){11}{\rule{0.786pt}{0.117pt}}
\multiput(552.00,349.17)(10.369,7.000){2}{\rule{0.393pt}{0.400pt}}
\multiput(564.00,357.59)(1.033,0.482){9}{\rule{0.900pt}{0.116pt}}
\multiput(564.00,356.17)(10.132,6.000){2}{\rule{0.450pt}{0.400pt}}
\multiput(576.00,363.59)(0.874,0.485){11}{\rule{0.786pt}{0.117pt}}
\multiput(576.00,362.17)(10.369,7.000){2}{\rule{0.393pt}{0.400pt}}
\multiput(588.00,370.59)(1.123,0.482){9}{\rule{0.967pt}{0.116pt}}
\multiput(588.00,369.17)(10.994,6.000){2}{\rule{0.483pt}{0.400pt}}
\multiput(601.00,376.59)(1.033,0.482){9}{\rule{0.900pt}{0.116pt}}
\multiput(601.00,375.17)(10.132,6.000){2}{\rule{0.450pt}{0.400pt}}
\multiput(613.00,382.59)(0.874,0.485){11}{\rule{0.786pt}{0.117pt}}
\multiput(613.00,381.17)(10.369,7.000){2}{\rule{0.393pt}{0.400pt}}
\multiput(625.00,389.59)(1.123,0.482){9}{\rule{0.967pt}{0.116pt}}
\multiput(625.00,388.17)(10.994,6.000){2}{\rule{0.483pt}{0.400pt}}
\multiput(638.00,395.59)(0.874,0.485){11}{\rule{0.786pt}{0.117pt}}
\multiput(638.00,394.17)(10.369,7.000){2}{\rule{0.393pt}{0.400pt}}
\multiput(650.00,402.59)(1.033,0.482){9}{\rule{0.900pt}{0.116pt}}
\multiput(650.00,401.17)(10.132,6.000){2}{\rule{0.450pt}{0.400pt}}
\multiput(662.00,408.59)(0.874,0.485){11}{\rule{0.786pt}{0.117pt}}
\multiput(662.00,407.17)(10.369,7.000){2}{\rule{0.393pt}{0.400pt}}
\multiput(674.00,415.59)(1.123,0.482){9}{\rule{0.967pt}{0.116pt}}
\multiput(674.00,414.17)(10.994,6.000){2}{\rule{0.483pt}{0.400pt}}
\multiput(687.00,421.59)(1.033,0.482){9}{\rule{0.900pt}{0.116pt}}
\multiput(687.00,420.17)(10.132,6.000){2}{\rule{0.450pt}{0.400pt}}
\multiput(699.00,427.59)(0.874,0.485){11}{\rule{0.786pt}{0.117pt}}
\multiput(699.00,426.17)(10.369,7.000){2}{\rule{0.393pt}{0.400pt}}
\multiput(711.00,434.59)(1.123,0.482){9}{\rule{0.967pt}{0.116pt}}
\multiput(711.00,433.17)(10.994,6.000){2}{\rule{0.483pt}{0.400pt}}
\multiput(724.00,440.59)(0.874,0.485){11}{\rule{0.786pt}{0.117pt}}
\multiput(724.00,439.17)(10.369,7.000){2}{\rule{0.393pt}{0.400pt}}
\multiput(736.00,447.59)(1.033,0.482){9}{\rule{0.900pt}{0.116pt}}
\multiput(736.00,446.17)(10.132,6.000){2}{\rule{0.450pt}{0.400pt}}
\multiput(748.00,453.59)(0.874,0.485){11}{\rule{0.786pt}{0.117pt}}
\multiput(748.00,452.17)(10.369,7.000){2}{\rule{0.393pt}{0.400pt}}
\multiput(760.00,460.59)(1.123,0.482){9}{\rule{0.967pt}{0.116pt}}
\multiput(760.00,459.17)(10.994,6.000){2}{\rule{0.483pt}{0.400pt}}
\multiput(773.00,466.59)(1.033,0.482){9}{\rule{0.900pt}{0.116pt}}
\multiput(773.00,465.17)(10.132,6.000){2}{\rule{0.450pt}{0.400pt}}
\multiput(785.00,472.59)(0.874,0.485){11}{\rule{0.786pt}{0.117pt}}
\multiput(785.00,471.17)(10.369,7.000){2}{\rule{0.393pt}{0.400pt}}
\multiput(797.00,479.59)(1.123,0.482){9}{\rule{0.967pt}{0.116pt}}
\multiput(797.00,478.17)(10.994,6.000){2}{\rule{0.483pt}{0.400pt}}
\multiput(810.00,485.59)(0.874,0.485){11}{\rule{0.786pt}{0.117pt}}
\multiput(810.00,484.17)(10.369,7.000){2}{\rule{0.393pt}{0.400pt}}
\multiput(822.00,492.59)(1.033,0.482){9}{\rule{0.900pt}{0.116pt}}
\multiput(822.00,491.17)(10.132,6.000){2}{\rule{0.450pt}{0.400pt}}
\multiput(834.00,498.59)(0.874,0.485){11}{\rule{0.786pt}{0.117pt}}
\multiput(834.00,497.17)(10.369,7.000){2}{\rule{0.393pt}{0.400pt}}
\multiput(846.00,505.59)(1.123,0.482){9}{\rule{0.967pt}{0.116pt}}
\multiput(846.00,504.17)(10.994,6.000){2}{\rule{0.483pt}{0.400pt}}
\multiput(859.00,511.59)(0.874,0.485){11}{\rule{0.786pt}{0.117pt}}
\multiput(859.00,510.17)(10.369,7.000){2}{\rule{0.393pt}{0.400pt}}
\multiput(871.00,518.59)(1.033,0.482){9}{\rule{0.900pt}{0.116pt}}
\multiput(871.00,517.17)(10.132,6.000){2}{\rule{0.450pt}{0.400pt}}
\multiput(883.00,524.59)(1.123,0.482){9}{\rule{0.967pt}{0.116pt}}
\multiput(883.00,523.17)(10.994,6.000){2}{\rule{0.483pt}{0.400pt}}
\multiput(896.00,530.59)(0.874,0.485){11}{\rule{0.786pt}{0.117pt}}
\multiput(896.00,529.17)(10.369,7.000){2}{\rule{0.393pt}{0.400pt}}
\multiput(908.00,537.59)(1.033,0.482){9}{\rule{0.900pt}{0.116pt}}
\multiput(908.00,536.17)(10.132,6.000){2}{\rule{0.450pt}{0.400pt}}
\multiput(920.00,543.59)(0.874,0.485){11}{\rule{0.786pt}{0.117pt}}
\multiput(920.00,542.17)(10.369,7.000){2}{\rule{0.393pt}{0.400pt}}
\multiput(932.00,550.59)(1.123,0.482){9}{\rule{0.967pt}{0.116pt}}
\multiput(932.00,549.17)(10.994,6.000){2}{\rule{0.483pt}{0.400pt}}
\multiput(945.00,556.59)(0.874,0.485){11}{\rule{0.786pt}{0.117pt}}
\multiput(945.00,555.17)(10.369,7.000){2}{\rule{0.393pt}{0.400pt}}
\multiput(957.00,563.59)(1.033,0.482){9}{\rule{0.900pt}{0.116pt}}
\multiput(957.00,562.17)(10.132,6.000){2}{\rule{0.450pt}{0.400pt}}
\multiput(969.00,569.59)(1.123,0.482){9}{\rule{0.967pt}{0.116pt}}
\multiput(969.00,568.17)(10.994,6.000){2}{\rule{0.483pt}{0.400pt}}
\multiput(982.00,575.59)(0.874,0.485){11}{\rule{0.786pt}{0.117pt}}
\multiput(982.00,574.17)(10.369,7.000){2}{\rule{0.393pt}{0.400pt}}
\multiput(994.00,582.59)(1.033,0.482){9}{\rule{0.900pt}{0.116pt}}
\multiput(994.00,581.17)(10.132,6.000){2}{\rule{0.450pt}{0.400pt}}
\multiput(1006.00,588.59)(0.874,0.485){11}{\rule{0.786pt}{0.117pt}}
\multiput(1006.00,587.17)(10.369,7.000){2}{\rule{0.393pt}{0.400pt}}
\multiput(1018.00,595.59)(1.123,0.482){9}{\rule{0.967pt}{0.116pt}}
\multiput(1018.00,594.17)(10.994,6.000){2}{\rule{0.483pt}{0.400pt}}
\multiput(1031.00,601.59)(0.874,0.485){11}{\rule{0.786pt}{0.117pt}}
\multiput(1031.00,600.17)(10.369,7.000){2}{\rule{0.393pt}{0.400pt}}
\multiput(1043.00,608.59)(1.033,0.482){9}{\rule{0.900pt}{0.116pt}}
\multiput(1043.00,607.17)(10.132,6.000){2}{\rule{0.450pt}{0.400pt}}
\multiput(1055.00,614.59)(1.123,0.482){9}{\rule{0.967pt}{0.116pt}}
\multiput(1055.00,613.17)(10.994,6.000){2}{\rule{0.483pt}{0.400pt}}
\multiput(1068.00,620.59)(0.874,0.485){11}{\rule{0.786pt}{0.117pt}}
\multiput(1068.00,619.17)(10.369,7.000){2}{\rule{0.393pt}{0.400pt}}
\multiput(1080.00,627.59)(1.033,0.482){9}{\rule{0.900pt}{0.116pt}}
\multiput(1080.00,626.17)(10.132,6.000){2}{\rule{0.450pt}{0.400pt}}
\multiput(1092.00,633.59)(0.874,0.485){11}{\rule{0.786pt}{0.117pt}}
\multiput(1092.00,632.17)(10.369,7.000){2}{\rule{0.393pt}{0.400pt}}
\multiput(1104.00,640.59)(1.123,0.482){9}{\rule{0.967pt}{0.116pt}}
\multiput(1104.00,639.17)(10.994,6.000){2}{\rule{0.483pt}{0.400pt}}
\multiput(1117.00,646.59)(0.874,0.485){11}{\rule{0.786pt}{0.117pt}}
\multiput(1117.00,645.17)(10.369,7.000){2}{\rule{0.393pt}{0.400pt}}
\multiput(1129.00,653.59)(1.033,0.482){9}{\rule{0.900pt}{0.116pt}}
\multiput(1129.00,652.17)(10.132,6.000){2}{\rule{0.450pt}{0.400pt}}
\multiput(1141.00,659.59)(1.033,0.482){9}{\rule{0.900pt}{0.116pt}}
\multiput(1141.00,658.17)(10.132,6.000){2}{\rule{0.450pt}{0.400pt}}
\multiput(1153.00,665.59)(0.950,0.485){11}{\rule{0.843pt}{0.117pt}}
\multiput(1153.00,664.17)(11.251,7.000){2}{\rule{0.421pt}{0.400pt}}
\multiput(1166.00,672.59)(1.033,0.482){9}{\rule{0.900pt}{0.116pt}}
\multiput(1166.00,671.17)(10.132,6.000){2}{\rule{0.450pt}{0.400pt}}
\multiput(1178.00,678.59)(0.874,0.485){11}{\rule{0.786pt}{0.117pt}}
\multiput(1178.00,677.17)(10.369,7.000){2}{\rule{0.393pt}{0.400pt}}
\multiput(1190.00,685.59)(1.123,0.482){9}{\rule{0.967pt}{0.116pt}}
\multiput(1190.00,684.17)(10.994,6.000){2}{\rule{0.483pt}{0.400pt}}
\multiput(1203.00,691.59)(0.874,0.485){11}{\rule{0.786pt}{0.117pt}}
\multiput(1203.00,690.17)(10.369,7.000){2}{\rule{0.393pt}{0.400pt}}
\multiput(1215.00,698.59)(1.033,0.482){9}{\rule{0.900pt}{0.116pt}}
\multiput(1215.00,697.17)(10.132,6.000){2}{\rule{0.450pt}{0.400pt}}
\multiput(1227.00,704.59)(1.033,0.482){9}{\rule{0.900pt}{0.116pt}}
\multiput(1227.00,703.17)(10.132,6.000){2}{\rule{0.450pt}{0.400pt}}
\multiput(1239.00,710.59)(0.950,0.485){11}{\rule{0.843pt}{0.117pt}}
\multiput(1239.00,709.17)(11.251,7.000){2}{\rule{0.421pt}{0.400pt}}
\multiput(1252.00,717.59)(1.033,0.482){9}{\rule{0.900pt}{0.116pt}}
\multiput(1252.00,716.17)(10.132,6.000){2}{\rule{0.450pt}{0.400pt}}
\multiput(1264.00,723.59)(0.874,0.485){11}{\rule{0.786pt}{0.117pt}}
\multiput(1264.00,722.17)(10.369,7.000){2}{\rule{0.393pt}{0.400pt}}
\multiput(1276.00,730.59)(1.123,0.482){9}{\rule{0.967pt}{0.116pt}}
\multiput(1276.00,729.17)(10.994,6.000){2}{\rule{0.483pt}{0.400pt}}
\multiput(1289.00,736.59)(0.874,0.485){11}{\rule{0.786pt}{0.117pt}}
\multiput(1289.00,735.17)(10.369,7.000){2}{\rule{0.393pt}{0.400pt}}
\multiput(1301.00,743.59)(1.033,0.482){9}{\rule{0.900pt}{0.116pt}}
\multiput(1301.00,742.17)(10.132,6.000){2}{\rule{0.450pt}{0.400pt}}
\multiput(1313.00,749.59)(1.033,0.482){9}{\rule{0.900pt}{0.116pt}}
\multiput(1313.00,748.17)(10.132,6.000){2}{\rule{0.450pt}{0.400pt}}
\multiput(1325.00,755.59)(0.950,0.485){11}{\rule{0.843pt}{0.117pt}}
\multiput(1325.00,754.17)(11.251,7.000){2}{\rule{0.421pt}{0.400pt}}
\multiput(1338.00,762.59)(1.033,0.482){9}{\rule{0.900pt}{0.116pt}}
\multiput(1338.00,761.17)(10.132,6.000){2}{\rule{0.450pt}{0.400pt}}
\multiput(1350.00,768.59)(0.874,0.485){11}{\rule{0.786pt}{0.117pt}}
\multiput(1350.00,767.17)(10.369,7.000){2}{\rule{0.393pt}{0.400pt}}
\multiput(1362.00,775.59)(1.123,0.482){9}{\rule{0.967pt}{0.116pt}}
\multiput(1362.00,774.17)(10.994,6.000){2}{\rule{0.483pt}{0.400pt}}
\multiput(1375.00,781.59)(0.874,0.485){11}{\rule{0.786pt}{0.117pt}}
\multiput(1375.00,780.17)(10.369,7.000){2}{\rule{0.393pt}{0.400pt}}
\multiput(1387.00,788.59)(1.033,0.482){9}{\rule{0.900pt}{0.116pt}}
\multiput(1387.00,787.17)(10.132,6.000){2}{\rule{0.450pt}{0.400pt}}
\multiput(1399.00,794.59)(1.033,0.482){9}{\rule{0.900pt}{0.116pt}}
\multiput(1399.00,793.17)(10.132,6.000){2}{\rule{0.450pt}{0.400pt}}
\multiput(1411.00,800.59)(0.950,0.485){11}{\rule{0.843pt}{0.117pt}}
\multiput(1411.00,799.17)(11.251,7.000){2}{\rule{0.421pt}{0.400pt}}
\multiput(1424.00,807.59)(1.033,0.482){9}{\rule{0.900pt}{0.116pt}}
\multiput(1424.00,806.17)(10.132,6.000){2}{\rule{0.450pt}{0.400pt}}
\end{picture}

\caption{Thermal diffusivity coefficient for different values of
the overlapping. Also shown (solid line) is the kinetic theory prediction.
}
\label{ds2}
\end{center}
\end{figure}

\begin{figure}[ht]
\begin{center}
\setlength{\unitlength}{0.240900pt}
\ifx\plotpoint\undefined\newsavebox{\plotpoint}\fi
\begin{picture}(1500,900)(0,0)
\font\gnuplot=cmr10 at 10pt
\gnuplot
\sbox{\plotpoint}{\rule[-0.200pt]{0.400pt}{0.400pt}}%
\put(220.0,113.0){\rule[-0.200pt]{0.400pt}{184.048pt}}
\put(220.0,113.0){\rule[-0.200pt]{4.818pt}{0.400pt}}
\put(198,113){\makebox(0,0)[r]{0.04}}
\put(1416.0,113.0){\rule[-0.200pt]{4.818pt}{0.400pt}}
\put(220.0,240.0){\rule[-0.200pt]{4.818pt}{0.400pt}}
\put(198,240){\makebox(0,0)[r]{0.05}}
\put(1416.0,240.0){\rule[-0.200pt]{4.818pt}{0.400pt}}
\put(220.0,368.0){\rule[-0.200pt]{4.818pt}{0.400pt}}
\put(198,368){\makebox(0,0)[r]{0.06}}
\put(1416.0,368.0){\rule[-0.200pt]{4.818pt}{0.400pt}}
\put(220.0,495.0){\rule[-0.200pt]{4.818pt}{0.400pt}}
\put(198,495){\makebox(0,0)[r]{0.07}}
\put(1416.0,495.0){\rule[-0.200pt]{4.818pt}{0.400pt}}
\put(220.0,622.0){\rule[-0.200pt]{4.818pt}{0.400pt}}
\put(198,622){\makebox(0,0)[r]{0.08}}
\put(1416.0,622.0){\rule[-0.200pt]{4.818pt}{0.400pt}}
\put(220.0,750.0){\rule[-0.200pt]{4.818pt}{0.400pt}}
\put(198,750){\makebox(0,0)[r]{0.09}}
\put(1416.0,750.0){\rule[-0.200pt]{4.818pt}{0.400pt}}
\put(220.0,877.0){\rule[-0.200pt]{4.818pt}{0.400pt}}
\put(198,877){\makebox(0,0)[r]{0.1}}
\put(1416.0,877.0){\rule[-0.200pt]{4.818pt}{0.400pt}}
\put(220.0,113.0){\rule[-0.200pt]{0.400pt}{4.818pt}}
\put(220,68){\makebox(0,0){0}}
\put(220.0,857.0){\rule[-0.200pt]{0.400pt}{4.818pt}}
\put(441.0,113.0){\rule[-0.200pt]{0.400pt}{4.818pt}}
\put(441,68){\makebox(0,0){2000}}
\put(441.0,857.0){\rule[-0.200pt]{0.400pt}{4.818pt}}
\put(662.0,113.0){\rule[-0.200pt]{0.400pt}{4.818pt}}
\put(662,68){\makebox(0,0){4000}}
\put(662.0,857.0){\rule[-0.200pt]{0.400pt}{4.818pt}}
\put(883.0,113.0){\rule[-0.200pt]{0.400pt}{4.818pt}}
\put(883,68){\makebox(0,0){6000}}
\put(883.0,857.0){\rule[-0.200pt]{0.400pt}{4.818pt}}
\put(1104.0,113.0){\rule[-0.200pt]{0.400pt}{4.818pt}}
\put(1104,68){\makebox(0,0){8000}}
\put(1104.0,857.0){\rule[-0.200pt]{0.400pt}{4.818pt}}
\put(1325.0,113.0){\rule[-0.200pt]{0.400pt}{4.818pt}}
\put(1325,68){\makebox(0,0){10000}}
\put(1325.0,857.0){\rule[-0.200pt]{0.400pt}{4.818pt}}
\put(220.0,113.0){\rule[-0.200pt]{292.934pt}{0.400pt}}
\put(1436.0,113.0){\rule[-0.200pt]{0.400pt}{184.048pt}}
\put(220.0,877.0){\rule[-0.200pt]{292.934pt}{0.400pt}}
\put(45,495){\makebox(0,0){$D$}}
\put(828,23){\makebox(0,0){$n$}}
\put(220.0,113.0){\rule[-0.200pt]{0.400pt}{184.048pt}}
\put(331,265){\raisebox{-.8pt}{\makebox(0,0){$\Diamond$}}}
\put(773,243){\raisebox{-.8pt}{\makebox(0,0){$\Diamond$}}}
\put(1325,247){\raisebox{-.8pt}{\makebox(0,0){$\Diamond$}}}
\put(331.0,213.0){\rule[-0.200pt]{0.400pt}{25.054pt}}
\put(321.0,213.0){\rule[-0.200pt]{4.818pt}{0.400pt}}
\put(321.0,317.0){\rule[-0.200pt]{4.818pt}{0.400pt}}
\put(773.0,224.0){\rule[-0.200pt]{0.400pt}{9.154pt}}
\put(763.0,224.0){\rule[-0.200pt]{4.818pt}{0.400pt}}
\put(763.0,262.0){\rule[-0.200pt]{4.818pt}{0.400pt}}
\put(1325.0,240.0){\rule[-0.200pt]{0.400pt}{3.132pt}}
\put(1315.0,240.0){\rule[-0.200pt]{4.818pt}{0.400pt}}
\put(1315.0,253.0){\rule[-0.200pt]{4.818pt}{0.400pt}}
\put(220,797){\usebox{\plotpoint}}
\put(220.0,797.0){\rule[-0.200pt]{292.934pt}{0.400pt}}
\put(220,247){\usebox{\plotpoint}}
\put(220.00,247.00){\usebox{\plotpoint}}
\put(240.76,247.00){\usebox{\plotpoint}}
\multiput(245,247)(20.756,0.000){0}{\usebox{\plotpoint}}
\put(261.51,247.00){\usebox{\plotpoint}}
\multiput(269,247)(20.756,0.000){0}{\usebox{\plotpoint}}
\put(282.27,247.00){\usebox{\plotpoint}}
\put(303.02,247.00){\usebox{\plotpoint}}
\multiput(306,247)(20.756,0.000){0}{\usebox{\plotpoint}}
\put(323.78,247.00){\usebox{\plotpoint}}
\multiput(331,247)(20.756,0.000){0}{\usebox{\plotpoint}}
\put(344.53,247.00){\usebox{\plotpoint}}
\put(365.29,247.00){\usebox{\plotpoint}}
\multiput(367,247)(20.756,0.000){0}{\usebox{\plotpoint}}
\put(386.04,247.00){\usebox{\plotpoint}}
\multiput(392,247)(20.756,0.000){0}{\usebox{\plotpoint}}
\put(406.80,247.00){\usebox{\plotpoint}}
\put(427.55,247.00){\usebox{\plotpoint}}
\multiput(429,247)(20.756,0.000){0}{\usebox{\plotpoint}}
\put(448.31,247.00){\usebox{\plotpoint}}
\multiput(453,247)(20.756,0.000){0}{\usebox{\plotpoint}}
\put(469.07,247.00){\usebox{\plotpoint}}
\put(489.82,247.00){\usebox{\plotpoint}}
\multiput(490,247)(20.756,0.000){0}{\usebox{\plotpoint}}
\put(510.58,247.00){\usebox{\plotpoint}}
\multiput(515,247)(20.756,0.000){0}{\usebox{\plotpoint}}
\put(531.33,247.00){\usebox{\plotpoint}}
\multiput(539,247)(20.756,0.000){0}{\usebox{\plotpoint}}
\put(552.09,247.00){\usebox{\plotpoint}}
\put(572.84,247.00){\usebox{\plotpoint}}
\multiput(576,247)(20.756,0.000){0}{\usebox{\plotpoint}}
\put(593.60,247.00){\usebox{\plotpoint}}
\multiput(601,247)(20.756,0.000){0}{\usebox{\plotpoint}}
\put(614.35,247.00){\usebox{\plotpoint}}
\put(635.11,247.00){\usebox{\plotpoint}}
\multiput(638,247)(20.756,0.000){0}{\usebox{\plotpoint}}
\put(655.87,247.00){\usebox{\plotpoint}}
\multiput(662,247)(20.756,0.000){0}{\usebox{\plotpoint}}
\put(676.62,247.00){\usebox{\plotpoint}}
\put(697.38,247.00){\usebox{\plotpoint}}
\multiput(699,247)(20.756,0.000){0}{\usebox{\plotpoint}}
\put(718.13,247.00){\usebox{\plotpoint}}
\multiput(724,247)(20.756,0.000){0}{\usebox{\plotpoint}}
\put(738.89,247.00){\usebox{\plotpoint}}
\put(759.64,247.00){\usebox{\plotpoint}}
\multiput(760,247)(20.756,0.000){0}{\usebox{\plotpoint}}
\put(780.40,247.00){\usebox{\plotpoint}}
\multiput(785,247)(20.756,0.000){0}{\usebox{\plotpoint}}
\put(801.15,247.00){\usebox{\plotpoint}}
\put(821.91,247.00){\usebox{\plotpoint}}
\multiput(822,247)(20.756,0.000){0}{\usebox{\plotpoint}}
\put(842.66,247.00){\usebox{\plotpoint}}
\multiput(846,247)(20.756,0.000){0}{\usebox{\plotpoint}}
\put(863.42,247.00){\usebox{\plotpoint}}
\multiput(871,247)(20.756,0.000){0}{\usebox{\plotpoint}}
\put(884.18,247.00){\usebox{\plotpoint}}
\put(904.93,247.00){\usebox{\plotpoint}}
\multiput(908,247)(20.756,0.000){0}{\usebox{\plotpoint}}
\put(925.69,247.00){\usebox{\plotpoint}}
\multiput(932,247)(20.756,0.000){0}{\usebox{\plotpoint}}
\put(946.44,247.00){\usebox{\plotpoint}}
\put(967.20,247.00){\usebox{\plotpoint}}
\multiput(969,247)(20.756,0.000){0}{\usebox{\plotpoint}}
\put(987.95,247.00){\usebox{\plotpoint}}
\multiput(994,247)(20.756,0.000){0}{\usebox{\plotpoint}}
\put(1008.71,247.00){\usebox{\plotpoint}}
\put(1029.46,247.00){\usebox{\plotpoint}}
\multiput(1031,247)(20.756,0.000){0}{\usebox{\plotpoint}}
\put(1050.22,247.00){\usebox{\plotpoint}}
\multiput(1055,247)(20.756,0.000){0}{\usebox{\plotpoint}}
\put(1070.98,247.00){\usebox{\plotpoint}}
\put(1091.73,247.00){\usebox{\plotpoint}}
\multiput(1092,247)(20.756,0.000){0}{\usebox{\plotpoint}}
\put(1112.49,247.00){\usebox{\plotpoint}}
\multiput(1117,247)(20.756,0.000){0}{\usebox{\plotpoint}}
\put(1133.24,247.00){\usebox{\plotpoint}}
\multiput(1141,247)(20.756,0.000){0}{\usebox{\plotpoint}}
\put(1154.00,247.00){\usebox{\plotpoint}}
\put(1174.75,247.00){\usebox{\plotpoint}}
\multiput(1178,247)(20.756,0.000){0}{\usebox{\plotpoint}}
\put(1195.51,247.00){\usebox{\plotpoint}}
\multiput(1203,247)(20.756,0.000){0}{\usebox{\plotpoint}}
\put(1216.26,247.00){\usebox{\plotpoint}}
\put(1237.02,247.00){\usebox{\plotpoint}}
\multiput(1239,247)(20.756,0.000){0}{\usebox{\plotpoint}}
\put(1257.77,247.00){\usebox{\plotpoint}}
\multiput(1264,247)(20.756,0.000){0}{\usebox{\plotpoint}}
\put(1278.53,247.00){\usebox{\plotpoint}}
\put(1299.29,247.00){\usebox{\plotpoint}}
\multiput(1301,247)(20.756,0.000){0}{\usebox{\plotpoint}}
\put(1320.04,247.00){\usebox{\plotpoint}}
\multiput(1325,247)(20.756,0.000){0}{\usebox{\plotpoint}}
\put(1340.80,247.00){\usebox{\plotpoint}}
\put(1361.55,247.00){\usebox{\plotpoint}}
\multiput(1362,247)(20.756,0.000){0}{\usebox{\plotpoint}}
\put(1382.31,247.00){\usebox{\plotpoint}}
\multiput(1387,247)(20.756,0.000){0}{\usebox{\plotpoint}}
\put(1403.06,247.00){\usebox{\plotpoint}}
\put(1423.82,247.00){\usebox{\plotpoint}}
\multiput(1424,247)(20.756,0.000){0}{\usebox{\plotpoint}}
\put(1436,247){\usebox{\plotpoint}}
\end{picture}

\caption{Thermal diffusivity coefficient for different values of
the density. Also shown (solid line) is the kinetic theory prediction
for this value of the overlapping $s=1.5$.
}
\label{dn}
\end{center}
\end{figure}
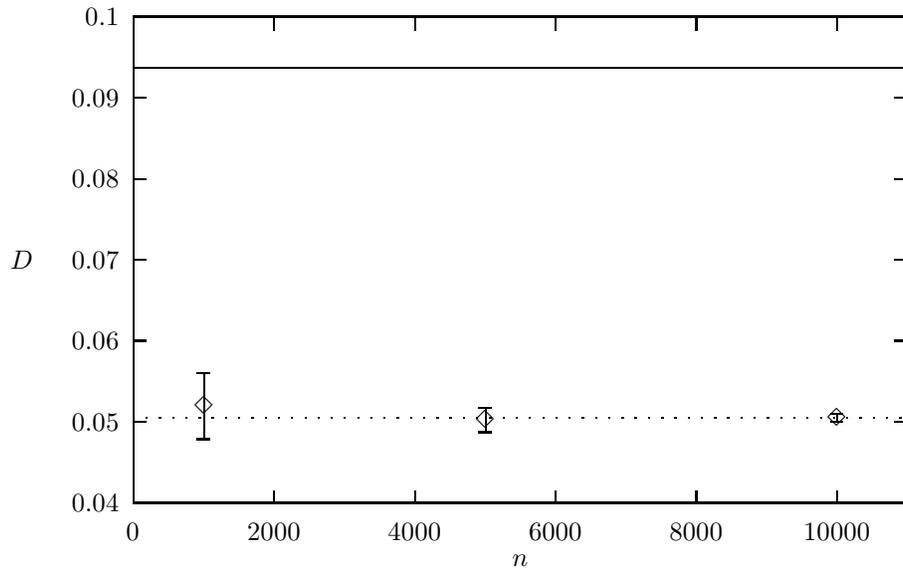


\section*{References}
\noindent

\begin{thebibliography}{000}
\bibitem{hoo92} P.J. Hoogerbrugge and J.M.V.A. Koelman,
Europhys. Lett. {\bf 19}, 155 (1992).

\bibitem{koe93} J.M.V.A. Koelman and P.J. Hoogerbrugge,
Europhys. Lett. {\bf 21}, 369 (1993).

\bibitem{boe97} E.S. Boek, P.V. Coveney, H.N.W. Lekkerkerker,
 and P. van der Schoot, Phys. Rev. E {\bf 55}, 3124 (1997).

\bibitem{boe97b} E.S. Boek, P.V. Coveney, and H.N.W. Lekkerkerker,
J. Phys.: Condens. Matter {\bf 8},  9509 (1997).

\bibitem{sch95} A.G. Schlijper, P.J. Hoogerbrugge, and C.W. Manke,
J. Rheol. {\bf 39}, 567 (1995).

\bibitem{cov97} P.V. Coveney and K. Novik, Phys. Rev. E
{\bf 54}, 5134 (1996).

\bibitem{esp95}
P. Espa{\~{n}}ol and P. Warren, Europhys. Lett. {\bf 30}, 191 (1995).

\bibitem{esp95a}
P. Espa{\~{n}}ol, Phys. Rev. E, {\bf 52}, 1734 (1995).

\bibitem{mar97} C. Marsh, G. Backx, and M.H. Ernst, Europhys. Lett.
{\bf 38}, 411 (1997).  C. Marsh, G. Backx, and M.H. Ernst,
 Phys. Rev. E  {\bf 56}, 1976 (1997).

\bibitem{esp97con} P. Espa\~nol, Europhys. Lett. {\bf 40}, 631 (1997).

\bibitem{bon97} J. Bonet Aval\'os and M. Mackie,
 Europhys. Lett. {\bf 40},  141 (1997).


\bibitem{lan59} L.D. Landau and E.M. Lifshitz, {\em Fluid Mechanics}
(Pergamon Press, 1959).

\bibitem{luc77} L.B. Lucy, Astron. J. {\bf 82}, 1013 (1977).

\bibitem{esp97sde} P. Espa\~nol, Physica  A {\bf 248}, 77 (1997).

\bibitem{all} M.P. Allen and D.J. Tildesley, {\em Computer Simulation
of Liquids} (Clarendon Press, Oxford, 1987).

\bibitem{preprint} M. Ripoll, M.H. Espa\~nol, and M.H. Ernst, preprint.

\bibitem{esp97fpm} P. Espa\~nol, Phys. Rev. E {\bf 57}, 2930 (1998).

\end{thebibliography}
\end{document}